\documentclass[traditabstract,article]{aa}
\pdfoutput=1
\usepackage{txfonts}
\usepackage{natbib}
\usepackage{graphicx}
\usepackage{afterpage}
\usepackage{lscape}
\usepackage{psfrag}
\usepackage{url}
\usepackage{microtype}
\usepackage{float}
\usepackage{afterpage}
\usepackage{longtable}
\usepackage{fixltx2e}
\usepackage[colorlinks=true, linkcolor=blue, citecolor=blue, urlcolor=blue]{hyperref}
\usepackage{breakurl}
\usepackage{xcolor}
\usepackage{flushend}
\usepackage[bottom]{footmisc}
\usepackage{lipsum}

\def\apjl{ApJL}

\def\aap{A\&A}
\def\apjs{ApJS}

\makeatletter
\renewcommand*{\@fnsymbol}[1]{$\star$}
\makeatother

\begin{document}
   \title{VALES: $\mathrm{II}$. The physical conditions of interstellar gas in normal \\ star-forming galaxies up to $z=0.2$ revealed by ALMA}

   \subtitle{ }
   
   \author{T.~M.~Hughes\inst{1}\thanks{email: thomas.hughes@uv.cl}, E.~Ibar\inst{1}, V.~Villanueva\inst{1}, M.~Aravena\inst{2}, M.~Baes\inst{3}, N.~Bourne\inst{4}, A.~Cooray\inst{5}, L.~Dunne\inst{4,6}, S.~Dye\inst{7},\\ S.~Eales\inst{6}, C.~Furlanetto\inst{7,8}, R.~Herrera-Camus\inst{9}, R.~J.~Ivison\inst{4,10}, E.~van~Kampen\inst{10}, M.~A.~Lara-L\'{o}pez\inst{11},\\S.~J.~Maddox\inst{4,6}, M.~J.~Micha{\l}owski\inst{4}, M.~W.~L.~Smith\inst{6}, E.~Valiante\inst{6}, P.~van~der~Werf\inst{12}, Y.~Q.~Xue\inst{13}}
	  
\institute{
     Instituto de F\'{i}sica y Astronom\'{i}a, Universidad de Valpara\'{i}so, Avda. Gran Breta\~{n}a 1111, Valpara\'{i}so, Chile
\and N\'{u}cleo de Astronom\'{i}a, Facultad de Ingenier\'{i}a, Universidad Diego Portales, Av. Ej\'{e}rcito 441, Santiago, Chile
\and Sterrenkundig Observatorium, Universiteit Gent, Krijgslaan 281-S9, Gent 9000, Belgium
\and Institute for Astronomy, University of Edinburgh, Royal Observatory, Edinburgh EH9 3HJ, UK
\and Department of Physics and Astronomy, University of California, Irvine, CA 92697, USA
\and School of Physics and Astronomy, Cardiff University, The Parade, Cardiff CF24 3AA, UK
\and School of Physics and Astronomy, University of Nottingham, University Park, Nottingham NG7 2RD, UK
\and CAPES Foundation, Ministry of Education of Brazil, Bras\'{i}lia/DF 70040-020, Brazil
\and Max-Planck-Institut für extraterrestrische Physik, Giessenbachstraße, 85748 Garching, Germany
\and European Southern Observatory, Karl-Schwarzschild-Strasse 2, 85748, Garching, Germany
\and Instituto de Astronom\'{i}a, Universidad Nacional Autonoma de M\'{e}xico, A.P. 70-264, 04510 M\'{e}xico, D.F., M\'{e}xico
\and Leiden Observatory, Leiden University, PO Box 9513, 2300 RA Leiden, The Netherlands
\and CAS Key Laboratory for Researches in Galaxies and Cosmology, Center for Astrophysics, Department of Astronomy, \\University of Science and Technology of China, Chinese Academy of Sciences, Hefei, Anhui 230026, China
}

   \date{Accepted for publication in A\&A.}

\newcommand{\hi}{H{\sc i}} 
\newcommand{\hii}{H{\sc ii}\ }
\newcommand{\cii}{[C{\sc ii}]}
\newcommand{\con}{$^{12}$CO(1--0)}
\newcommand{\co}{CO(1--0)}
\newcommand{\ci}{[C{\sc i}]}
\newcommand{\lco}{$L_{\mathrm{CO}}$}
\newcommand{\lcii}{$L_{\mathrm{[C{\textsc{ii}}]}}$}
\newcommand{\oi}{[O{\sc i}]}
\newcommand{\oii}{[O{\sc ii}]}
\newcommand{\oiii}{[O{\sc iii}]}
\newcommand{\oiv}{[O{\sc iv}]}
\newcommand{\nii}{[N{\sc ii}]}
\newcommand{\niii}{[N{\sc iii}]}
\newcommand{\ha}{H{\sc $\alpha$}}
\newcommand{\hd}{H{\sc $\delta$}}
\newcommand{\hg}{H{\sc $\gamma$}}
\newcommand{\hb}{H{\sc $\beta$}}
\newcommand{\kms}{km~s$^{-1}$\ }
\newcommand{\sdust}{$\Sigma_{\mathrm{dust}}$}
\newcommand{\sgas}{$\Sigma_{\mathrm{gas}}$}
\newcommand{\shii}{$\Sigma_{\mathrm{H}_{2}}$} 
\newcommand{\shi}{$\Sigma_{\mathrm{H}\tiny{\textsc{i}}}$} 
\newcommand{\ssfr}{$\Sigma_{\mathrm{SFR}}$}
\newcommand{\tir}{$F_{\mathrm{TIR}}$}
\newcommand{\fir}{$F_{\mathrm{FIR}}$}
\newcommand{\ciitir}{[C{\sc ii}]/$F_{\mathrm{TIR}}$}
\newcommand{\ciifir}{[C{\sc ii}]/$F_{\mathrm{FIR}}$}
\newcommand{\ciioitir}{([C{\sc ii}]+[O{\sc i}]63)/$F_{\mathrm{TIR}}$}
\newcommand{\ciioipah}{([C{\sc ii}]+[O{\sc i}]63)/$F_{\mathrm{PAH}}$}
\newcommand{\ciioialttir}{([C{\sc ii}]+[O{\sc i}]145)/$F_{\mathrm{TIR}}$}
\newcommand{\ltir}{$L_{\mathrm{TIR}}$}
\newcommand{\lfir}{$L_{\mathrm{FIR}}$}
\newcommand{\ciiltir}{[C{\sc ii}]/$L_{\mathrm{TIR}}$}
\newcommand{\ciilfir}{[C{\sc ii}]/$L_{\mathrm{FIR}}$}
\newcommand{\ciioiltir}{([C{\sc ii}]+[O{\sc i}]63)/$L_{\mathrm{TIR}}$}
\newcommand{\ciioilpah}{([C{\sc ii}]+[O{\sc i}]63)/$L_{\mathrm{PAH}}$}
\newcommand{\ciioi}{[C{\sc ii}]/[O{\sc i}]63}
\newcommand{\ciioialt}{[C{\sc ii}]/[O{\sc i}]145}
\newcommand{\ciico}{[C{\sc ii}]/CO(1--0)}
\newcommand{\ciipdr}{[C{\sc ii}]$_{\mathrm{PDR}}$}

\newcommand{\ciipdrob}{${\mathrm{[C\textsc{ii}]}}^{\,}_{\tiny\textsc{PDR}}$} 
\newcommand{\ciipdrhi}{${\mathrm{[C\textsc{ii}]}}^{\tiny 70\%}_{\tiny\textsc{PDR}}$} 
\newcommand{\ciipdrlo}{${\mathrm{[C\textsc{ii}]}}^{\tiny 50\%}_{\tiny\textsc{PDR}}$}

\newcommand{\ciiion}{${\mathrm{[C\textsc{ii}]}}^{\,}_{\tiny\textsc{ionised}}$} 
\newcommand{\ciisynion}{${\mathrm{[C\textsc{ii}]}}^{\tiny 24\,\mu\mathrm{m}}_{\tiny\textsc{ionised}}$} 
\newcommand{\ciiobsion}{${\mathrm{[C\textsc{ii}]}}^{\mathrm{\tiny[N\textsc{ii}]205}}_{\tiny\textsc{ionised}}$}

  \abstract{We use new Band-3 \co \ observations taken with the Atacama Large Millimeter/submillimeter Array (ALMA) to study the physical conditions in the interstellar gas of a sample of 27 dusty main-sequence star-forming galaxies at $0.03<z<0.2$ present in the Valpara\'{i}so ALMA Line Emission Survey (VALES). The sample is drawn from far-IR bright galaxies over $\sim$160 deg$^{2}$ in the \textit{Herschel} Astrophysical Terahertz Large Area Survey (\textit{H}-ATLAS), which is covered by high-quality ancillary data including \textit{Herschel} \cii \ 158~$\mu$m spectroscopy and far-infrared (FIR) photometry. The \cii \ and \co \ lines are both detected at $>5\sigma$ in 26 sources. We find an average \cii \ to CO(1--0) luminosity ratio of $3500\pm1200$ for our sample that is consistent with previous studies. Using the \cii , \co \ and FIR measurements as diagnostics of the physical conditions of the interstellar medium, we compare these observations to the predictions of a photodissociation region (PDR) model to determine the gas density, surface temperature, pressure, and the strength of the incident far-ultraviolet (FUV) radiation field, $G_{0}$, normalised to the Habing Field. The majority of our sample exhibit hydrogen densities of $4 < \log n/\mathrm{cm}^{3} < 5.5$ and experience an incident FUV radiation field with strengths of $2 < \log G_0 < 3$ when adopting standard adjustments. A comparison to galaxy samples at different redshifts indicates that the average strength of the FUV radiation field appears constant up to redshift $z\sim6.4$, yet the neutral gas density increases as a function of redshift by a factor of $\sim$100 from $z=0$ to $z=0.2$ that persists regardless of various adjustments to our observable quantities. Whilst this evolution could provide an explanation for the observed evolution of the star formation rate density with cosmic time, the result most likely arises from a combination of observational biases when using different suites of emission lines as diagnostic tracers of PDR gas.}

\keywords{galaxies: high-redshift -- galaxies: ISM -- infrared: galaxies  -- submillimeter: galaxies -- ISM: lines and bands } 

\authorrunning{T. M. Hughes et al.}
\titlerunning{Physical conditions of PDR gas in star-forming galaxies}
\maketitle

\section{Introduction}

The cosmic star formation rate density ($\rho_\mathrm{SFR}$) has declined by a factor of 20 since an observed peak at $z\sim$2.5 (\citealp*{hopkins2006}; \citealp*{madau2014}), and it remains unknown whether this is due to the exhaustion of the galactic interstellar medium (ISM), a reduction in the accretion of the pristine intergalactic medium, or a decline in the efficiency in the conversion of gas to stars. One approach towards disentangling the physical processes contributing to the decline in the overall $\rho_\mathrm{SFR}$ requires the characterisation of the content and physical conditions of the interstellar gas in galaxies at all redshifts. 

Probing the physical conditions of the ISM requires observations of emission lines, such as the far-infrared fine-structure lines of \cii \ 158~$\mu$m, \nii \ 122, 205~$\mu$m, \oi \ 63, 145~$\mu$m, and \oiii \ 88~$\mu$m lines, that play a crucial role in the thermal balance of the gas, and the rotational transitions of carbon monoxide (CO). In particular, the \cii ~158~$\mu$m line ($\nu_{\text{rest}}=1900.54\,$GHz), which originates from the $^{2}$P$_{3/2}$ $\rightarrow$ $^2$P$_{1/2}$ transition of the ground state of singly ionized carbon, typically has a luminosity of 0.1-1\% of the far-infrared luminosity in normal star-forming galaxies, thus making it one of the dominant cooling lines (e.g. \citealp*{dalgarno1972}; \citealp{crawford1985}; \citealp{stacey1991}). The \cii \ line emission comes from both neutral and ionised gas, as the low ionization potential of atomic carbon means C$^{+}$ can be produced from far-ultraviolet (FUV) photons with energies greater than just 11.26~eV (c.f. hydrogen's ionization potential of 13.6~eV). In star-forming galaxies, the majority of the line emission ($\sim$70\%; e.g. \citealp{stacey1991}; \citealp{stacey2010a}) is shown to arise from photodissociation regions (PDRs), with the remaining fraction coming from star-forming \hii regions, lower density warm gas and diffuse \hi \ clouds, X-ray dominated regions (XDRs) and cosmic ray dominated regions (CRDRs). Deeper within the PDR regions, C$^{+}$ becomes converted to CO -- a standard tracer of the molecular gas content. The \cii /\co \ ratio is mostly dependent on the C$^{+}$ column density and surface temperature, which both decrease with stronger gas shielding (\citealp{kaufman1999}). These lines are thus useful for constraining the ISM conditions in galaxies (see e.g. \citealp*{tielens1985}; \citealp{wolfire1990}).

Recent advancements in space- and ground-based facilities for observing these emission lines at far-infrared and sub-millimetre wavelengths are rapidly expanding our knowledge of the ISM in both nearby and high redshift galaxies (see e.g. \citealp*{solomon2005}; \citealp*{carilli2013}). The emission of \cii \ and other fine-structure lines has
been observed in low-$z$ galaxy samples with airborne- or space-based observatories, such as the \emph{Kuiper} Airborne Observatory (KAO; e.g. \citealp{stacey1991}; \citealp{madden1993}) and the Infrared Space Observatory (ISO; e.g. \citealp{hunter2001}; \citealp{malhotra2001}; \citealp{brauher2008}). The \textit{Herschel} Space Observatory \citep{pilbratt2010} with the PACS \citep{poglitsch2010} and SPIRE \citep{griffin2010} instruments was capable of observing both the FIR cooling lines and FIR/submm spectral energy distribution at unprecedented resolution, enabling the study of gas heating and cooling (via the \ciiltir \ or \ciioiltir \ ratios) on galactic and spatially-resolved, sub-kiloparsec scales (see e.g. \citealp{croxall2012}; \citealp{lebouteiller2012}; \citealp{parkin2013}; \citealp{hughes2015}). 

At higher redshifts, studies of the ISM primarily rely on observations of the \cii \ line and the rotational transitions of CO as diagnostics of the physical conditions. However, sources at $z>1$ avoid strong atmospheric absorption at submillimetre wavelengths and can be observed with ground-based instrumentation, such as e.g. the Northern Extended Millimetre Array (NOEMA), the Atacama Pathfinder EXperiment (APEX), and the Atacama Large Millimeter/submillimeter Array (ALMA). Numerous studies over the past decade (see e.g. \citealp{gullberg2015}, and references therein) report the detection of \cii \ emission in high-$z$ sources that are classified as possible active galactic nuclei (AGN) hosts (e.g. \citealp{maiolino2005}; \citealp{stacey2010a}; \citealp{wang2013}) or starburst galaxies (see e.g. \citealp{ivison2010}; \citealp{hailey2010}; \citealp{stacey2010a}; \citealp{cox2011}; \citealp{swinbank2012}; \citealp{riechers2013}; \citealp{magdis2014}; \citealp{brisbin2015}). A significant fraction of these detections also has \co \ observations, like in the sample of gravitationally-lensed, dusty star-forming galaxies in the redshift range $z\sim$2.1--5.7 (\citealp{gullberg2015}; \citealp{aravena2016b}) discovered by the South Pole Telescope (SPT; \citealp{carlstrom2011}). The increasing number of observations available for systems over a wide redshift range means we can begin to characterise the ISM physical conditions at various epochs.

The physical properties of the gaseous components of the ISM may be determined by comparing the observed ratio of the \cii \ 158~$\mu$m and \co \ line emission to the predictions of a PDR model. There are numerous PDR models available for determining the gas density, temperature and strength of the FUV radiation field (see \citealp{rollig2007} for a discussion, and references within). One of the most commonly used PDR models is that of \citet{tielens1985}, which characterises the physical conditions in a semi-infinite, plane-parallel slab PDR by two free variables: the hydrogen nucleus density, $n$, and the strength of the FUV (6 $< h\nu <$ 13.6~eV) radiation field, $G_0$, which is in units of the Habing Field (\citealp{habing1968}). The model has since been updated by \citet{wolfire1990}, \cite{hollenbach1991} and \citet{kaufman1999,kaufman2006}. Predictions from PDR models have been compared to \textit{Herschel} observations of both Galactic PDRs and nearby galaxies. For example, \citet{croxall2012} studied a late-type spiral, NGC~1097, and a Seyfert 1 galaxy, NGC~4559, finding $50 \le G_{0} \le 1000$ varying with $10^{2.5}\,\mathrm{cm}^{-3} \le n \le 10^{3}\,\mathrm{cm}^{-3}$ across both discs. Most recently, \citet{parkin2013} examined the $n$ and $G_0$ in various regions of M51; the hydrogen density and FUV radiation peak in the nucleus and similarly decline in both the spiral arm and interarm regions, suggesting similar physical conditions in clouds in these environments (see also \citealp{parkin2014}; \citealp{hughes2015}). \cite{stacey2010a} posit that the observed \lcii /\lfir \ and \lcii /\lco \ luminosity ratios suggest the gas density and FUV radiation field in their sample galaxies at $z\sim1-2$ are similar to the physical conditions in local starburst systems, which is also supported by the observations of the SPT sample (\citealp{gullberg2015}).

In this paper, we study the physical conditions in the interstellar gas for a sample of 27 dusty galaxies at $0.03<z<0.2$ selected from the Valpara\'{i}so ALMA Line Emission Survey (VALES; \citealp{villanuevaprep}), which are characterised as normal star-forming galaxies (see Fig. 1 of \citealp{villanuevaprep}) generally lying on or slightly above the main sequence (see e.g. \citealp{elbaz2011}). We use new ALMA Band-3 \co \ observations combined with \textit{Herschel}-PACS \cii \ 158~$\mu$m line emission data and far-infrared (FIR) luminosities determined with photometry from \textit{Herschel}-SPIRE and other facilities (\citealp{ibar2015}), to investigate the physical properties of the interstellar gas in these galaxies by using the PDR model of \citet{kaufman1999, kaufman2006}. We compare the physical conditions in our VALES sample to similar studies at low- and high-$z$. Our paper is structured as follows: in \hyperref[sec:sampledata]{Sec.~\ref*{sec:sampledata}}, we describe our sample, observations and data reduction methodology. In \hyperref[sec:ciicoratio]{Sec.~\ref*{sec:ciicoratio}} and \hyperref[sec:pdrmodelling]{Sec.~\ref*{sec:pdrmodelling}}, we describe the characteristics of the gas and compare our observations to theoretical PDR models. Finally, \hyperref[sec:discussion]{Sec.~\ref*{sec:discussion}} and ~\hyperref[sec:conclusions]{Sec.~\ref*{sec:conclusions}} present our discussion and conclusions. Throughout this paper, we adopt a $\Lambda$CDM cosmology with $H_0=70$\,km\,s$^{-1}$\,Mpc$^{-1}$, $\Omega_{\rm M}=0.27$ and $\Omega_{\Lambda}=0.73$.

\setlength{\tabcolsep}{4pt}
\renewcommand{\arraystretch}{1.2}
\begin{table*}[t]
\small
 \centering
  \caption{Properties of the targets analysed in this work. 
    Column 1: Galaxy ID from \textit{H}-ATLAS DR1 (see \citealp{valiante2016});
    Column 2 and 3: Right ascension and declination (J2000);
    Column 4: GAMA spectroscopic redshift;
    Column 5: Luminosity distance;
    Column 6: Stellar mass from \citet{ibar2015};
    Column 7: FIR (8-1000$\mu$m) luminosity;
    Column 8: \cii ~158\,$\mu$m line flux density from \citet{ibar2015};
    Column 9: \cii ~158\,$\mu$m line luminosity; 
    Column 10: \co \ line flux density from \citet{villanuevaprep};
    Column 11: \co \ line luminosity.\vspace{-0.2in}
}\label{tab:summary}
\begin{center}
  \begin{tabular}{l c c c c c c c c c c}
  \hline
\hline
Galaxy ID & RA    & Dec   & $z_\mathrm{spec}$ & $D_\mathrm{L}$ & $\log M_{\star}$ & $\log L_{\mathrm{FIR}}$ & $S_{\mathrm{[CII]}}\Delta \mathrm{v}$ & $L_{\mathrm{[CII]}}$                 & $S_{\mathrm{CO}}\Delta \mathrm{v}$ & $L_{\mathrm{CO}}$             \\
     & hms   & dms   &     &  Mpc           & M$_{\odot}$                   & L$_{\odot}$             & Jy km s$^{-1}$               & $\times$10$^{8}$L$_{\odot}$          & Jy km s$^{-1}$            & $\times$10$^{5}$ L$_{\odot}$  \\
 (1)    & (2)  &  (3)   &  (4)   &  (5)           & (6)                   & (7)             & (8)   & (9)       & (10)  &   (11)  \\     
\hline
G09.DR1.12  & 09:09:49 & +01:48:47  &  0.182  &  886.7  &  11.25 $\pm$ 0.12  &  11.84 $\pm$ 0.02  &  1052 $\pm$ 51  &  13.81 $\pm$ 0.68  &  8.48 $\pm$ 0.58  &  6.76 $\pm$ 0.46  \\ 
G09.DR1.20  & 09:12:05 & +00:26:55  &  0.055  &  244.3  &  10.46 $\pm$ 0.12  &  11.09 $\pm$ 0.01  &  1291 $\pm$ 43  &  1.45 $\pm$ 0.05  &  13.64 $\pm$ 0.84  &  0.93 $\pm$ 0.06  \\ 
G09.DR1.24  & 08:36:01 & +00:26:17  &  0.033  &  146.7  &  10.55 $\pm$ 0.11  &  10.31 $\pm$ 0.02  &  2373 $\pm$ 44  &  0.97 $\pm$ 0.02  &  20.80 $\pm$ 2.40  &  0.52 $\pm$ 0.06  \\ 
G09.DR1.32  & 08:57:48 & +00:46:41  &  0.072  &  325.9  &  10.58 $\pm$ 0.13  &  11.27 $\pm$ 0.01  &  2398 $\pm$ 43  &  4.69 $\pm$ 0.09  &  11.52 $\pm$ 0.58  &  1.37 $\pm$ 0.07  \\ 
G09.DR1.37  & 08:54:50 & +02:12:08  &  0.059  &  262.3  &  10.89 $\pm$ 0.11  &  10.70 $\pm$ 0.02  &  1794 $\pm$ 58  &  2.30 $\pm$ 0.08  &  12.87 $\pm$ 1.23  &  1.00 $\pm$ 0.10  \\ 
G09.DR1.43  & 09:00:05 & +00:04:46  &  0.054  &  241.5  &  10.81 $\pm$ 0.11  &  10.57 $\pm$ 0.02  &  1697 $\pm$ 53  &  1.86 $\pm$ 0.06  &  11.44 $\pm$ 1.64  &  0.76 $\pm$ 0.11  \\ 
G09.DR1.47  & 08:44:28 & +02:03:50  &  0.026  &  111.5  &  10.29 $\pm$ 0.11  &  10.25 $\pm$ 0.01  &  1331 $\pm$ 33  &  0.33 $\pm$ 0.01  &  14.00 $\pm$ 1.18  &  0.20 $\pm$ 0.02  \\ 
G09.DR1.49  & 08:53:46 & +00:12:52  &  0.051  &  225.6  &  10.28 $\pm$ 0.13  &  10.71 $\pm$ 0.01  &  2274 $\pm$ 41  &  2.18 $\pm$ 0.04  &  6.48 $\pm$ 0.13  &  0.38 $\pm$ 0.01  \\ 
G09.DR1.53  & 08:58:35 & +01:31:49  &  0.107  &  496.5  &  11.07 $\pm$ 0.12  &  11.22 $\pm$ 0.01  &  1201 $\pm$ 47  &  5.30 $\pm$ 0.21  &  9.64 $\pm$ 0.92  &  2.57 $\pm$ 0.25  \\ 
G09.DR1.56  & 08:51:11 & +01:30:06  &  0.060  &  267.2  &  10.63 $\pm$ 0.12  &  10.72 $\pm$ 0.02  &  1997 $\pm$ 55  &  2.66 $\pm$ 0.07  &  8.44 $\pm$ 1.52  &  0.68 $\pm$ 0.12  \\ 
G09.DR1.60  & 09:05:32 & +02:02:22  &  0.052  &  232.2  &  10.60 $\pm$ 0.12  &  10.69 $\pm$ 0.02  &  1832 $\pm$ 39  &  1.86 $\pm$ 0.04  &  19.20 $\pm$ 4.08  &  1.18 $\pm$ 0.25  \\ 
G09.DR1.61  & 08:58:28 & +00:38:13  &  0.053  &  234.5  &  10.75 $\pm$ 0.12  &  10.44 $\pm$ 0.02  &  907 $\pm$ 32  &  0.94 $\pm$ 0.03  &  3.44 $\pm$ 0.42  &  0.22 $\pm$ 0.03  \\ 
G09.DR1.62  & 08:46:30 & +00:50:55  &  0.133  &  625.7  &  10.73 $\pm$ 0.11  &  11.51 $\pm$ 0.02  &  909 $\pm$ 91  &  6.22 $\pm$ 0.63  &  5.54 $\pm$ 0.50  &  2.29 $\pm$ 0.21  \\ 
G09.DR1.72  & 08:44:28 & +02:06:59  &  0.079  &  358.8  &  10.59 $\pm$ 0.13  &  11.01 $\pm$ 0.03  &  1917 $\pm$ 58  &  4.51 $\pm$ 0.14  &  13.56 $\pm$ 1.78  &  1.94 $\pm$ 0.25  \\ 
G09.DR1.80  & 08:43:50 & +00:55:34  &  0.073  &  331.5  &  10.63 $\pm$ 0.12  &  11.03 $\pm$ 0.01  &  849 $\pm$ 46  &  1.70 $\pm$ 0.09  &  0.98 $\pm$ 0.22  &  0.12 $\pm$ 0.03  \\ 
G09.DR1.85  & 08:37:45 & -00:51:41  &  0.031  &  134.9  &  10.35 $\pm$ 0.12  &  10.13 $\pm$ 0.03  &  2090 $\pm$ 35  &  0.73 $\pm$ 0.01  &  6.72 $\pm$ 1.44  &  0.14 $\pm$ 0.03  \\ 
G09.DR1.87  & 08:52:34 & +01:34:19  &  0.195  &  958.2  &  10.59 $\pm$ 0.10  &  11.92 $\pm$ 0.01  &  $<$750  &  $<$11.38  &  10.75 $\pm$ 0.07  &  9.90 $\pm$ 0.06  \\ 
G09.DR1.99  & 09:07:50 & +01:01:41  &  0.128  &  604.0  &  10.51 $\pm$ 0.13  &  11.70 $\pm$ 0.01  &  1467 $\pm$ 62  &  9.33 $\pm$ 0.40  &  6.84 $\pm$ 0.58  &  2.65 $\pm$ 0.22  \\ 
G09.DR1.113 & 08:38:31 & +00:00:44  &  0.078  &  356.0  &  10.55 $\pm$ 0.13  &  11.15 $\pm$ 0.01  &  1052 $\pm$ 42  &  2.43 $\pm$ 0.10  &  8.78 $\pm$ 0.68  &  1.24 $\pm$ 0.10  \\ 
G09.DR1.125 & 08:53:40 & +01:33:48  &  0.041  &  182.2  &  10.48 $\pm$ 0.12  &  10.28 $\pm$ 0.03  &  1509 $\pm$ 31  &  0.95 $\pm$ 0.02  &  5.34 $\pm$ 2.10  &  0.20 $\pm$ 0.08  \\ 
G09.DR1.127 & 08:43:05 & +01:08:55  &  0.078  &  354.3  &  10.35 $\pm$ 0.14  &  11.05 $\pm$ 0.03  &  $<$543  &  $<$1.25  &  5.90 $\pm$ 0.60  &  0.82 $\pm$ 0.08  \\ 
G09.DR1.159 & 08:54:05 & +01:11:30  &  0.044  &  196.3  &  10.13 $\pm$ 0.13  &  10.54 $\pm$ 0.02  &  2349 $\pm$ 42  &  1.71 $\pm$ 0.03  &  4.80 $\pm$ 1.25  &  0.21 $\pm$ 0.06  \\ 
G09.DR1.179 & 08:49:07 & -00:51:38  &  0.070  &  316.5  &  10.48 $\pm$ 0.11  &  11.18 $\pm$ 0.01  &  1235 $\pm$ 46  &  2.28 $\pm$ 0.09  &  12.32 $\pm$ 0.98  &  1.38 $\pm$ 0.11  \\ 
G09.DR1.185 & 08:53:56 & +00:12:55  &  0.051  &  227.5  &  10.26 $\pm$ 0.13  &  10.33 $\pm$ 0.03  &  1439 $\pm$ 38  &  1.41 $\pm$ 0.04  &  9.54 $\pm$ 1.86  &  0.56 $\pm$ 0.11  \\ 
G09.DR1.276 & 08:51:12 & +01:03:42  &  0.027  &  117.3  &  9.94 $\pm$ 0.11  &  10.20 $\pm$ 0.01  &  901 $\pm$ 34  &  0.24 $\pm$ 0.01  &  6.30 $\pm$ 0.87  &  0.10 $\pm$ 0.01  \\ 
G09.DR1.294 & 08:42:17 & +02:12:23  &  0.096  &  443.3  &  10.53 $\pm$ 0.11  &  10.93 $\pm$ 0.04  &  648 $\pm$ 32  &  2.28 $\pm$ 0.11  &  5.44 $\pm$ 0.64  &  1.17 $\pm$ 0.14  \\ 
G09.DR1.328 & 08:41:39 & +01:53:46  &  0.074  &  334.8  &  10.54 $\pm$ 0.11  &  10.98 $\pm$ 0.01  &  880 $\pm$ 48  &  1.81 $\pm$ 0.10  & $<$1.30  &  $<$0.02  \\          
\hline
\end{tabular}
\end{center}
\end{table*}

\section{The sample and data}\label{sec:sampledata}

Our sample of galaxies is drawn from a \textit{Herschel} programme capable of providing a sufficient number of far-IR bright galaxies over $\sim$600 deg$^{2}$: the \textit{Herschel} Astrophysical Terahertz Large Area Survey (\citealp{eales2010}; \citealp{valiante2016}; \citealp{bourne2016}). In addition to a wealth of high-quality ancillary data, a significant sample have both \textit{Herschel}-PACS \cii \ 158~$\mu$m (\citealp{ibar2015}) and our follow-up ALMA \co \ emission line observations from VALES (\citealp{villanuevaprep}). From the three equatorial fields (totaling $\sim$160 deg$^{2}$) covered by $H$-ATLAS, galaxies were selected based on the following criteria: (1) a flux limit of $S_{160 \mu\mathrm{m}} >$ 150 mJy, i.e. near the peak of the SED of a typical local star-forming galaxy; (2) no neighbours with $S_{160 \mu\mathrm{m}} > 160$~mJy ($3 \sigma$) within 2 arcmin from their centroids; (3) an unambiguous identification (\textsc{reliability} $>$0.8, \citealp{bourne2016}) in the Sloan Digital Sky Survey (SDSS DR7; \citealp{abazajian2009}); (4) a Petrosian SDSS $r$-band radius $<$15$\arcsec$, i.e. smaller than the PACS spectroscopic field of view; (5) high-quality spectroscopic redshifts (\textsc{zqual} $>$ 3) from the Galaxy and Mass Assembly survey (GAMA; \citealp{driver2009,driver2011}; \citealp{liske2015}); and (6) a redshift between 0.02 $<z<$ 0.2 (median of 0.05), beyond which the \cii \ emission becomes redshifted to the edge of the PACS spectrometer 160~$\mu$m band. After applying these criteria, 324 galaxies remain to comprise a statistically-significant sample spanning a wide range of optical morphological types and IR luminosities. Of these, 27 objects have observations of both the \cii \ and \co \ emission line observations that form the focus of our study.

\subsection{Herschel-PACS \cii \ 158 micron line observations}

The \cii \ 158~$\mu$m line observations of our sample are presented in \citet{ibar2015}. The end of the \textit{Herschel} mission meant only 28 galaxies -- all located in the GAMA 9h field -- of the parent sample could be observed during our \textit{Herschel}-PACS \cii \ spectroscopic campaign. Their selection arises purely on the basis of scheduling efficiency, so this sample is representative of the original sample albeit smaller in number (see Fig.~1 of \citealp{ibar2015}). We first briefly summarise their properties.

The redshifted \cii ~158\,$\mu$m line emission was observed with the PACS first order ({\sc r1}) filter covering $\sim$\,2$\,\mu$m of bandwidth in a 47\,arcsec $\times$ 47\,arcsec field of view in a single pointing (pointed-mode). The central spaxel of the 5$\times$5 spaxal array captured the majority of the line emission for each target. Data cubes comprising $5\times 5$ spaxels $\times\, n_{\rm chan}$ rebinned spectral channels were generated from calibrated PACS level-2 data products processed with SPG v12.1.0, with an effective spectral resolution of $\sim$\,190--240\,km\,s$^{-1}$. The \cii ~158~$\mu$m line flux density was determined from the weighted sum (aided by the instrumental noise cube) of the central $3\times3$\,spaxels, via the simultaneous fitting of a linear background slope and a Gaussian to the spectra. Uncertainties at the 1-$\sigma$ level were derived for the line parameters with a Monte-Carlo realisation ($1000\times$), randomly varying the signal per spectral channel using the instrumental error cube. These \cii \ line flux measurements are presented in \hyperref[tab:summary]{Table~\ref*{tab:summary}}, yet we refer the reader to \citet{ibar2015} for more details.

\subsection{ALMA \co \ line observations}

Despite the fact that ALMA is not an ideal telescope for creating galaxy surveys, we are using a novel observational approach to target large numbers of galaxies at low/intermediate redshifts to create a reference sample for interpreting observations of the high redshift Universe. Since our sample is drawn from the GAMA fields that are each $\sim 4^\circ\times14^\circ$ in size, providing large numbers of galaxies at similar redshifts, we can minimise the number of spectral tunings needed to observe all sources independently by setting the source redshifts to zero and fixing the spectral windows (SPW) manually in order to cover the widest possible spectral (thus redshift) range. The central frequency position of each SPW is then optimized to cover the redshifted \co \ line for the maximum number of sources, therefore significantly minimising overheads. Using this observing strategy, we obtained observations targeting the \co \ line for 67 galaxies during cycle-1 and -2 (project 2013.1.00530.S; P.I.:\ E.\ Ibar). Whereas \citet{villanuevaprep} present the observations, data reduction and analysis in detail for the complete sample, in this paper we focus solely on galaxies with the cycle-2 CO(1–0) and {\it Herschel} \cii \ data. Here, we briefly summarise these observations and data reduction steps. 

The cycle-2 observations taken in Band-3 covered the \mbox{$^{12}$CO(1--0)} emission line down to $2$\,mJy\,beam$^{-1}$ at $30$ km s$^{-1}$ for a sample of 27 sources at $0.03 < z < 0.2$. All observations were reduced homogeneously within the \textit{Common Astronomy Software Applications}\footnote{\url{http://casa.nrao.edu/index.shtml}} (CASA; \citealp{mcmullin2007}) using a common pipeline, developed from standard pipelines, for calibration, concatenation and imaging. The bandpass calibrators for our sources were J0750+1231, J0739+0137 and J0909+0121, flux calibrators were Callisto, Mars, Ceres and Titan, and the phase calibrators were J0909+0121 and J0901-0037. The optimal image resolution was that which provides the highest number of non-cleaned point-like $>5 \sigma$ sources when varying the cube spectral resolution from 20 to 100~km~s$^{-1}$ in steps of 10~km~s$^{-1}$ with the {\sc clean} task; cubes without detections were set at 100~km~s$^{-1}$ channel width. The final cubes, of 256$\times$256 pixels in spatial size, were corrected for the primary beam, manually cleaned to a threshold of 3$\sigma$ and created using natural weighting. The barycenter velocity reference was set as the optical spectroscopic redshift ($z_\mathrm{spec}$) of each source.

For measuring the \co \ line emission, we first perform a Gaussian fit to the spectra to identify the central frequency, $\nu_{\mathrm{obs}}$, and $\nu_{\mathrm{FWHM}}$ of the line. The velocity integrated \co \ flux densities ($S_{\mathrm{CO}} \Delta \mathrm{v}$ in units of Jy~km~s$^{-1}$) were then obtained by collapsing the data cubes between $\nu_{\mathrm{obs}}-\nu_{\mathrm{FWHM}}$ and $\nu_{\mathrm{obs}}+\nu_{\mathrm{FWHM}}$, and fitting these cubes with a Gaussian. We detect $>95$\% (26 of 27) of the targets with a $>5 \sigma$ peak line detection. We note that our selection criteria includes targets with SDSS \textit{r}-band radii smaller than 15$\arcsec$, in order to obtain reliable \textit{Herschel}-PACS observations (\citealp{ibar2015}). Considering that the maximum recoverable angular scales of ALMA in Band-3 are approximately 25$\arcsec$ (with a 60$\arcsec$ primary beam), these detections do not suffer from any missing flux. For the upper–limit of source G09.DR1.328, we collapse the 100~km~s$^{-1}$ spectral resolution whilst setting $\nu_{\mathrm{FWHM}}$ = 250~km~s$^{-1}$ and adopt the limit as 5$\times$ the measured RMS. Our \co \ line flux measurements are presented in \hyperref[tab:summary]{Table~\ref*{tab:summary}} and we refer the reader to \citet{villanuevaprep} for more details. We calculate the CO line luminosity, \lco , in units of L$_{\odot}$ following
\begin{equation}
  L_{\mathrm{CO}} = 1.04 \times 10^{-3} S_{\mathrm{CO}}\Delta \mathrm{v}\, \nu_{\mathrm{rest}}(1+z)^{-1}D_{\mathrm{L}}^{2}\,, 
\end{equation}
where $\nu_{\mathrm{rest}}$ is the rest frequency in GHz and $D_{\mathrm{L}}$ is the luminosity distance in Mpc (from Eqn. 1 of \citealp*{solomon2005}).

\subsection{Ancillary data for far-infrared luminosities}

We use previously published values of the far-infrared (FIR) luminosity, $L_{\mathrm{FIR}}$, that were measured for each galaxy by fitting the SED constructed from {\it Herschel} PACS and SPIRE, {\it WISE}-22\,$\mu$m and {\it IRAS} photometry (see \citealp{ibar2013,ibar2015}). In brief, each rest-frame SED is fit with a modified black body that is forced to follow a power law at the high-frequency end of the spectrum and the flux of the best-fitting SED is integrated between $8\,  \mu$m and $1000\,\mu$m, i.e., 
\begin{equation}
  L_{\mathrm{FIR}}(8-1000\,\mu{\rm m}) =
  4\pi\,D_{\rm L}^2(z)\,\int_{\nu_{\rm 1}}^{\nu_{\rm 2}}S_\nu\,d\nu
\end{equation}
to obtain the dust temperature ($T_{\rm d}$), the dust emissivity index ($\beta$), the mid-IR slope ($\alpha_{\operatorname{mid-IR}}$), and the normalisation which provides the total FIR luminosity (rest-frame 8--1000\,$\mu$m). These \lfir \ estimates and their uncertainties, obtained from randomly varying the broadband photometry within their uncertainties in a Monte-Carlo simulation ($100\times$), are listed in \hyperref[tab:summary]{Table~\ref*{tab:summary}}, whereas the complete set of derived parameters may be found in Table 3 of \citet{ibar2015}. 

We note the fact that previous studies use differing definitions of the total infrared band (TIR, e.g. 3--1100\,$\mu$m; \citealp{parkin2013,parkin2014}) and also the FIR bands that are commonly used as proxies of the total infrared emission, such as from $42\,  \mu$m to $122\,\mu$m (e.g \citealp{helou1988}; \citealp{dale2001}; \citealp{stacey2010a}), or from $40\,  \mu$m to $500\,\mu$m (e.g. \citealp{graciacarpio2008}; \citealp{gullberg2015}). These quantities are related by $F_{\mathrm{TIR}}$(3--1100\,$\mu$m) $\approx$ 2$\times F_{\mathrm{FIR}}$(42--122\,$\mu$m) in \citet{dale2001} and $F_{\mathrm{TIR}}$(8--1000\,$\mu$m) $\approx$ 1.3$\times F_{\mathrm{FIR}}$(40--500\,$\mu$m) in \citet{graciacarpio2008}. The difference in the IR luminosities obtained over 3--1100\,$\mu$m and 8--1000\,$\mu$m is typically less than 15\% (e.g \citealp{rosario2016}), thus only a small fraction of the TIR emission. Therefore, the FIR luminosity as defined in this present work (8--1000\,$\mu$m) is a good proxy for the TIR emission and is approximately equivalent to the bolometric far-infrared flux of the PDR model (see e.g. \citealp{parkin2013,parkin2014}), which we discuss shortly.  

\begin{figure}
\begin{center}
\includegraphics[width=0.99\columnwidth]{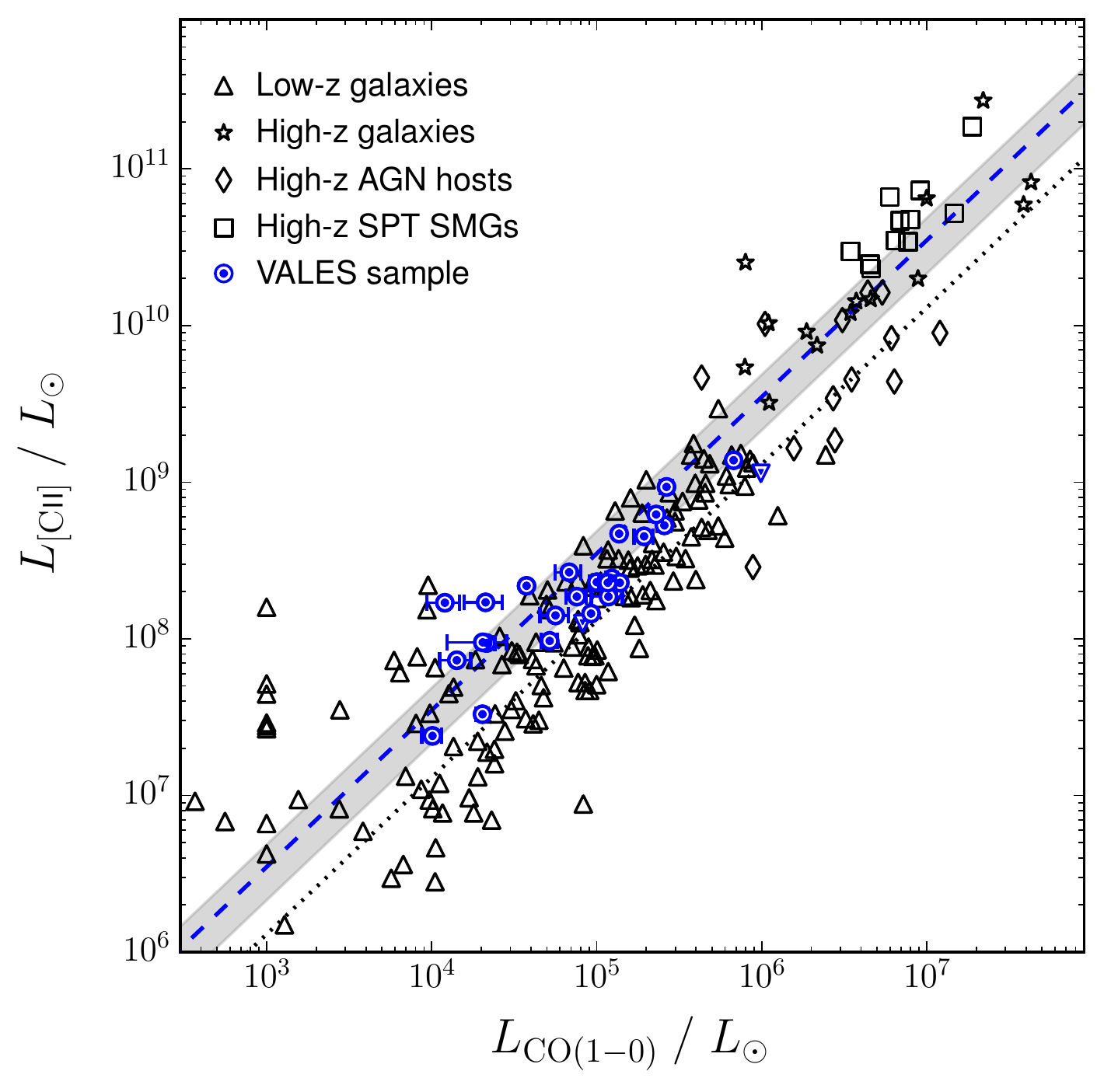}
\end{center}
\vspace{-0.5cm}
\caption{The \cii \ luminosity versus the \co \ luminosity for the VALES sample at $0.03<z<0.2$ superimposed on the low- and high-$z$ samples taken from Fig. 7 of \citealp{gullberg2015}, with symbols as specified in the legend. Downward triangles represent 5$\sigma$ upper limits. The grey shaded region denotes the 1-$\sigma$ spread in the mean observed \lcii /\lco \ ratio (3500$\pm$1200, blue dashed line) of our sample, which we compare to the mean ratios of the low-z sample (1300$\pm$440, black dotted line).}\label{fig:lciilco}
\end{figure}

\section{The \cii \ -- CO luminosity correlation}\label{sec:ciicoratio}

Before deriving the physical conditions via PDR modelling, we first perform a brief sanity check on the observations of our VALES sample by examining the well-studied \cii --CO luminosity correlation (see e.g. \citealp{crawford1985}). Previous studies have shown that a variety of sources, from normal star-forming galaxies to starbursts and AGN hosts, follow the correlation up to $z\sim 6$ with an observed \lcii / \lco \ ratio ranging from 1300 to 6300 with a typical median of 4400 (\citealp{crawford1985}; \citealp{stacey1991,stacey2010a}; \citealp{swinbank2012}). In \hyperref[fig:lciilco]{Fig.~\ref*{fig:lciilco}}, we plot the \cii \ luminosity versus the \co \ luminosity and superimpose our \textit{H}-ATLAS-based sample at $0.03<z<0.2$ onto the observations of samples at higher and lower redshifts presented by \citet[][see their Fig. 7]{gullberg2015}. We find the average \lcii /\lco \ ratio for our sample is 3500 (median of $\sim$2400) with a standard deviation of 1200, slightly higher than the average value (1300$\pm$440) for the low-$z$ normal star-forming galaxies but much lower than that of the high-$z$ SPT sample (5200$\pm$1800; \citealp{gullberg2015}).

\subsection{The origin of the line emission}

Following the reasoning and methodology of \citet{gullberg2015}, we use the \lcii /\lco \ ratio to constrain the optical depths and excitation temperatures of the lines from a comparison of the source functions. The \lcii /\lco \ ratio is then given by
 \newline
\begin{align}\label{eqn:ciicorat}
\begin{split}
\frac{L_{\mathrm{[C{\textsc{ii}}]}}}{L_{\mathrm{CO(1-0)}}} = & \left( \frac{\nu_{\mathrm{[C{\textsc{ii}}]}}}{\nu_{\mathrm{CO(1-0)}}} \right)^{3} \times \left( \frac{\Delta \nu_{\mathrm{[C{\textsc{ii}}]}}}{\Delta \nu_{\mathrm{CO(1-0)}}} \right)\times  \frac{1-e^{-\tau_{\mathrm{[C{\textsc{ii}}]}}}}{1-e^{-\tau_{\mathrm{CO(1-0)}}}} \\
 & \times \frac{e^{h\nu_{\mathrm{CO(1-0)}}/kT_{\mathrm{ex,CO(1-0)}}}-1}{e^{h\nu_{[C{\textsc{ii}}]}/kT_{\mathrm{ex,[C{\textsc{ii}}]}}}-1} \, \mathrm{,}
\end{split}
\end{align}
 \newline
\noindent under the assumption that the \cii \ and  \co \ filling factors are equal. Knowing the \lcii /\lco \ ratio leaves the two excitation temperatures ($T_{\mathrm{ex,[C{\textsc{ii}}]}}$, $T_{\mathrm{ex,CO(1-0)}}$) and two opacities ($\tau_{\mathrm{[C{\textsc{ii}}]}}$, $\tau_{\mathrm{CO(1-0)}}$) as free parameters, which we can then vary to match the observations. 

We first consider the scenario when the excitation temperatures are equal, $T_{\mathrm{ex,[C{\textsc{ii}}]}}=T_{\mathrm{ex,CO(1-0)}}$. In this case, \hyperref[eqn:ciicorat]{Eqn.~\ref*{eqn:ciicorat}} will under predict the observed \lcii /\lco \ ratio by an order of magnitude when the \cii \ is optically thin ($\tau_{\mathrm{[C{\textsc{ii}}]}}=0.1$) and CO is optically thick ($\tau_{\mathrm{CO(1-0)}}=1$), but will over predict the ratio by an order of magnitude when \cii \ is optically thick ($\tau_{\mathrm{[C{\textsc{ii}}]}}=1$) and CO is optically thin ($\tau_{\mathrm{CO(1-0)}}=0.1$). The observed ratio is only reproducible when the optical depths are the same ($\tau_{\mathrm{[C{\textsc{ii}}]}}=\tau_{\mathrm{CO(1-0)}}$) but requires equal excitation temperatures of $>50$~K and, given that the CO is usually optically thick (e.g., from the $^{12}$CO/$^{13}$CO line ratios, see \citealp{hughesprep}), this implies that the \cii \ line would also be approaching the optically thick regime. However, such a scenario where both \cii \ and CO lines are optically thick with equal excitation temperatures over 50 K is not a viable solution given the observational evidence to the contrary. Firstly, \cii \ optical depths are consistently of the order unity or less for even very bright Galactic star formation regions (\citealp{stacey1991b}; \citealp*{boreiko1996}; \citealp{graf2012}; \citealp{ossenkopf2013}). Secondly, although the observed \cii /\oi line ratios (see e.g. \citealp{stacey1983b}; \citealp{lord1996}; \citealp{brauher2008}) and measurements using velocity-resolved peak \cii \ antenna temperatures (\citealp{graf2012}; \citealp{ossenkopf2013}) yield \cii \ excitation temperatures that are far in excess of 50 K, CO excitation temperatures on galactic scales are typically less than 50 K (e.g. \citealp{gullberg2015}).

\begin{figure}
\begin{center}
\includegraphics[width=0.99\columnwidth]{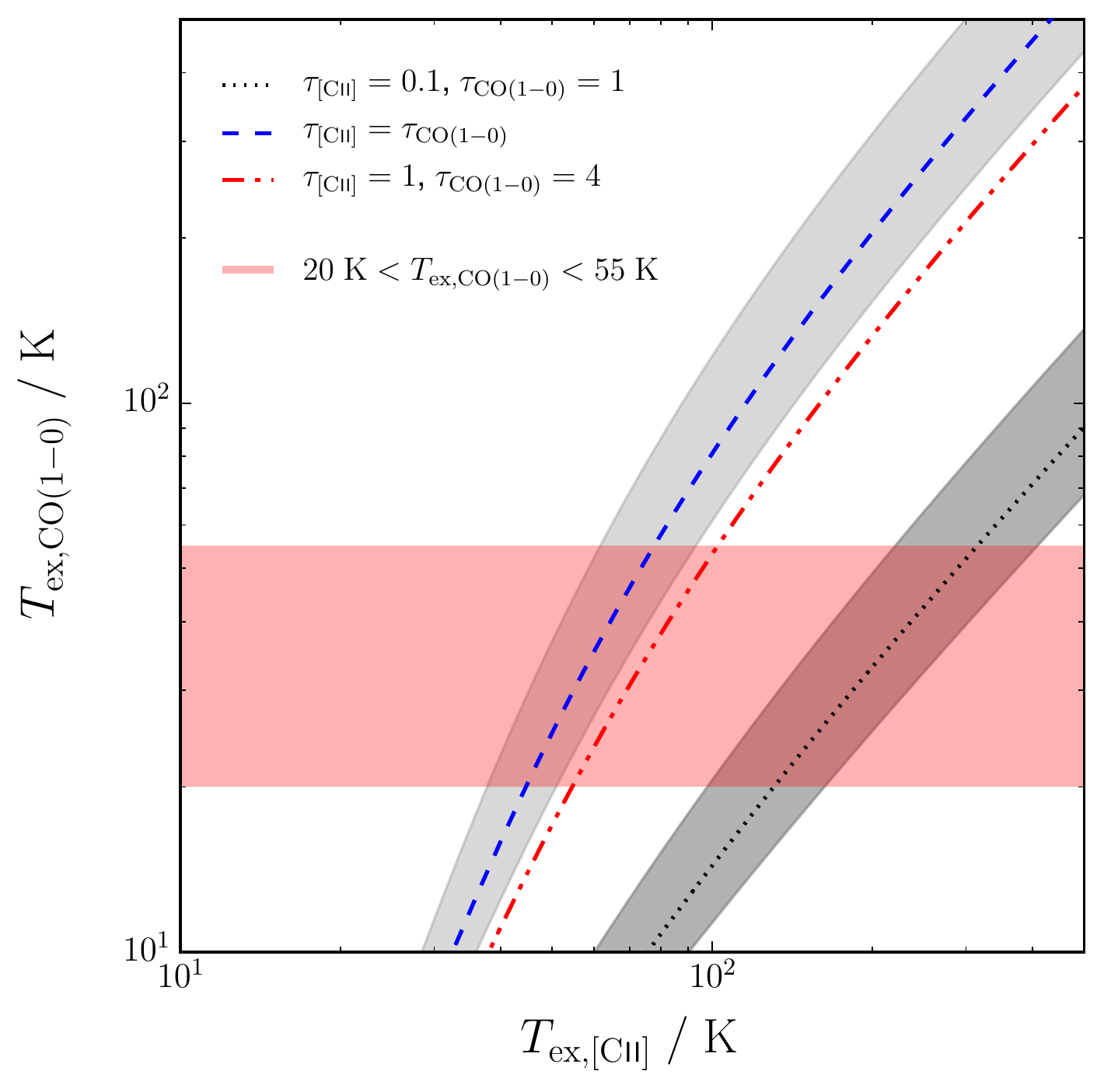}
\end{center}
\vspace{-0.5cm}
\caption{The \co \ versus \cii \ excitation temperatures determined via the ratio of the source functions (\hyperref[eqn:ciicorat]{Eqn.~\ref*{eqn:ciicorat}}) assuming various line opacities, as specified in the legend, and adopting the observed \lcii /\lco \ ratio of 3500. The light and dark grey shaded regions denote denotes the 1-$\sigma$ spread in the mean observed ratio, whereas the red shaded region represents the CO excitation temperatures of $20~\mathrm{K}<T_{\mathrm{ex,CO(1-0)}} < 55~\mathrm{K}$. Combined with the results of PDR modelling and previous observational evidence in the literature (\hyperref[sec:insights]{Sec.~\ref*{sec:insights}}), the best scenario to explain the observed \lcii /\lco \ ratios is gas where $T_{\mathrm{ex,[C{\textsc{ii}}]}} > T_{\mathrm{ex,CO(1-0)}}$ and with \cii \ is optically thin and the CO is optically thick ($\tau_{\mathrm{[C{\textsc{ii}}]}}=0.1$, $\tau_{\mathrm{CO(1-0)}}=1$; black dotted line).}\label{fig:temperature}
\end{figure}

The alternative scenario to consider is when the two excitation temperatures are different. We examine the relationship between $T_{\mathrm{ex,[C{\textsc{ii}}]}}$ and $T_{\mathrm{ex,CO(1-0)}}$ for our observed \lcii /\lco \ ratio considering various optical depths of the two lines (see \hyperref[fig:temperature]{Fig.~\ref*{fig:temperature}}), finding that $T_{\mathrm{ex,[C{\textsc{ii}}]}} > T_{\mathrm{ex,CO(1-0)}}$ in all cases and that we cannot exclude both \cii \ and CO \ being optically thick ($\tau_{\mathrm{[C{\textsc{ii}}]}}=1$, $\tau_{\mathrm{CO(1-0)}}>1$). In addition, we can fix the CO excitation temperature as in \citet{weiss2013} and \citet{gullberg2015} by making the assumption that the CO traces molecular gas, within which dust is thermalised, such that the dust temperatures obtained for our VALES sample ($20~\mathrm{K}< T_{\mathrm{d}} < 55~\mathrm{K}$; \citealp{ibar2015}) are representative of $T_{\mathrm{ex,CO(1-0)}}$. From \hyperref[fig:temperature]{Fig.~\ref*{fig:temperature}}, the fixed range of $T_{\mathrm{ex,CO(1-0)}}$ implies $T_{\mathrm{ex,[C{\textsc{ii}}]}} \approx$ 35--90~K for $\tau_{\mathrm{[C{\textsc{ii}}]}}=\tau_{\mathrm{CO(1-0)}}$ and $T_{\mathrm{ex,[C{\textsc{ii}}]}} \approx$ 45--110~K for $\tau_{\mathrm{[C{\textsc{ii}}]}}=1$ with $\tau_{\mathrm{CO(1-0)}}=4$. However, we also predict that our \lcii /\lco \ ratio can arise from gas with $T_{\mathrm{ex,[C{\textsc{ii}}]}} \approx$ 100--300~K with optically thin \cii \ ($\tau_{\mathrm{[C{\textsc{ii}}]}}=0.1$) and optically thick CO ($\tau_{\mathrm{CO(1-0)}}=1$). When the gas density is greater than the \cii \ critical density, the excitation temperature becomes equivalent to the gas temperature and, as we present in the following section (\hyperref[sec:insights]{Sec.~\ref*{sec:insights}}), the PDR models support this latter scenario. 

To summarise, we conclude that the best scenario to explain the \lcii /\lco \ ratios observed in our VALES sample is where the \cii \ and CO emission originates from gas with $T_{\mathrm{ex,[C{\textsc{ii}}]}} > T_{\mathrm{ex,CO(1-0)}}$, where the \cii \ is optically thin and the CO is optically thick.

\section{Results from PDR modelling}\label{sec:pdrmodelling}

We now compare our observations to the PDR model of \citet{kaufman1999,kaufman2006}, an updated version of the model of \citet*{tielens1985}. The model treats PDR regions as homogeneous infinite plane slabs of hydrogen with physical conditions characterised by the hydrogen nuclei density, $n$, and the strength of the incident FUV radiation field, $G_{0}$, which is normalised to the \citeauthor{habing1968} Field in units of $1.6 \times 10^{-3}$~erg~cm$^{-2}$~s$^{-1}$ (\citealp{habing1968}). The gas is collisionally heated via the ejection of photoelectrons from dust grains and PAH molecules by FUV photons, and gas cooling from line emission is predicted by simultaneously solving the chemical and energy equilibrium in the slab. For a given a set of observations of spectral line intensities, the corresponding $G_{0}$ and $n$ values predicted by the PDR model are available online\footnote{The PDR Toolbox is available online at \url{http://dustem.astro.umd.edu}} via the `Photo Dissociation Region Toolbox' \citep*[PDRT,][]{pound2008}, where the models cover a density range of $10^{1}\, \le n \le 10^{7}\,\mathrm{cm}^{-3}$ and a FUV radiation field strength range of $10^{-0.5} \le G_{0} \le 10^{6.5}$.  In the following, the \cii \ and \co \ observations are compared to the PDR model grid lines from the \citet{kaufman1999} diagnostic plots, where we must assume that each emission component -- the \cii \ line emission, the \co \ emission, and the far-IR continuum -- originates from a single PDR component in our sources. 

\begin{figure}
\begin{center}
\includegraphics[width=0.99\columnwidth]{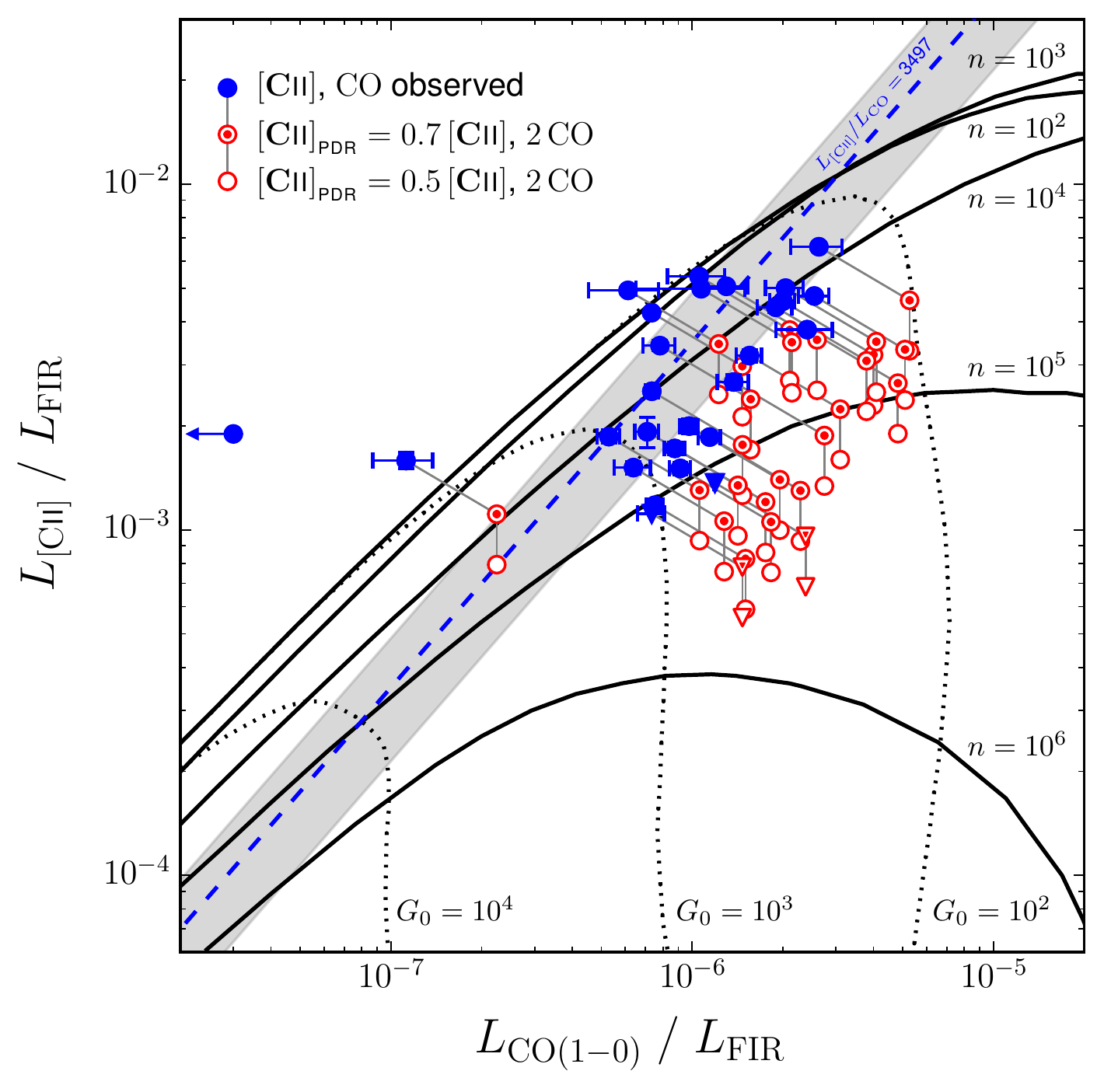}
\end{center}
\vspace{-0.5cm}
\caption{Diagnostic diagram of the observed \lcii/\lfir \ ratio versus \lco/\lfir \ ratio for our sample. We superimpose our observations onto the grid of constant hydrogen nuclei density, log $n$ (black solid lines), and FUV radiation field strength, log $G_{0}$ (black dotted lines), determined from the PDR model of \citet{kaufman1999,kaufman2006}. We present our unadjusted observations (blue solid circles) and the observations including the adjustments applied to the \cii \ and \co \ emission as described in \hyperref[sec:adjustments]{Sec.~\ref*{sec:adjustments}}, and compare the effects of our two estimations of the fraction of the \cii \ emission arising from ionised gas, where \cii $_{\mathrm{PDR}}$ is 70\% (red semi-open circles) and 50\% (red open circles) of the total \cii \ emission. Downward triangles represent 5$\sigma$ upper limits. Two sources lie above the $\log n = 2$ contour, i.e. outside the model parameter space, and the blue arrow points towards G09.DR1.328, which remains outside the plot regardless of adjustments. The grey shaded region denotes the 1-$\sigma$ spread in the mean observed \lcii /\lco \ ratio (blue dashed line). This figure is adapted from \citet{stacey2010a} and \citet{hailey2010}.}\label{fig:pdrdiag}
\end{figure}

In \hyperref[fig:pdrdiag]{Fig.~\ref*{fig:pdrdiag}}, we superimpose the observed (i.e. unadjusted) \lcii \ versus the \lco \ line luminosities for our sample on the PDR model grid lines of constant log ($n/\mathrm{cm}^{-3}$) and log $G_{0}$ constructed from the \citet{kaufman1999,kaufman2006} diagnostic plots, adapted from figures in \citet{stacey2010a} and \citet{hailey2010}. The majority of our observations lie within the parameter space covered by the PDR model and exhibit moderate FUV radiation field strengths ($2.0 < \log G_0 < 3.0$) and moderate hydrogen densities ($2 < \log n/\mathrm{cm}^{-3} < 4.5$). We stress, however, that there is much uncertainty in the lower limit of the latter parameter, owing to the degeneracy in the parameter space between densities of $\log n/\mathrm{cm}^{3} = 2$ and $3$. In addition, three of our galaxies fall outside of the \lcii/\lfir \ versus the \lco/\lfir \ parameter space defined by the PDR model - the closest contours of constant $\log n$ and $\log G_{0}$ are those of the lowest density and weakest field strength, respectively. We remove galaxy G09.DR1.328 from the remainder of the analysis, since the estimated \lco/\lfir \ of 2$\times 10^{-8}$ (from 5$\times$ the RMS) places the galaxy far outside the parameter space explored in \hyperref[fig:pdrdiag]{Fig.~\ref*{fig:pdrdiag}}. Before continuing further, several adjustments to our observations are necessary in order to facilitate a proper comparison with the PDR model.

\setlength{\tabcolsep}{6pt}
\renewcommand{\arraystretch}{1.2}
\begin{table*}[t]
\small
 \centering
  \caption{The average gas physical conditions - hydrogen nuclei density ($n$), FUV radiation field strength ($G_0$), temperture ($T$) and pressure ($P$) - derived from the comparison of our observations and the \citet{kaufman1999,kaufman2006} PDR model, when adopting various adjustments to the observed quantities (see \hyperref[sec:adjustments]{Sec.~\ref*{sec:adjustments}}) and considering each galaxy as a single PDR component. We note that pressure is $P = nT$ with units of K cm$^{-3}$.\vspace{-0.2in}}\label{tab:params}
\begin{center}
\begin{tabular}{l c c c c c c c c c c c c c c}
\hline 
\hline
\multicolumn{1}{l}{ } & \multicolumn{4}{c}{\cii , CO~ observed} & & \multicolumn{4}{c}{\ciipdrhi \ and 2CO } & & \multicolumn{4}{c}{\ciipdrlo \ and 2CO} \\ \cline{2-5} \cline{7-10} \cline{12-15}
Target  &  $\log n/\mathrm{cm}^{3}$ &$\log G_0$& $T$/K & $P$/10$^{6}$ & & $\log n/\mathrm{cm}^{3}$ &$\log G_0$& $T$/K & $P$/10$^{6}$ & & $\log n/\mathrm{cm}^{3}$ &$\log G_0$& $T$/K & $P$/10$^{6}$\\
\hline
G09.DR1.12   &  4.6   &  2.9   &  124  &  4.9   &   &  5.3  &  2.8  &  255  &  50.9   &   &  5.6   &  2.8  &  568  &  226.1  \\ 
G09.DR1.20   &  5.0   &  3.0   &  186  &  18.6  &   &  5.6  &  2.9  &  568  &  226.1  &   &  5.8   &  2.9  &  1190 &  750.8 \\ 
G09.DR1.24   &  4.2   &  2.5   &  172  &  2.7   &   &  4.8  &  2.1  &  73   &  4.6    &   &  5.0   &  2.2  &  78   &  7.8  \\ 
G09.DR1.32   &  4.0   &  2.9   &  209  &  2.1   &   &  5.0  &  2.8  &  140  &  14.0   &   &  5.3   &  2.9  &  255  &  50.9  \\ 
G09.DR1.37   &  4.1   &  2.6   &  172  &  2.2   &   &  4.7  &  2.3  &  88   &  4.4    &   &  5.0   &  2.4  &  91   &  9.1  \\ 
G09.DR1.43   &  4.0   &  2.6   &  172  &  1.7   &   &  4.7  &  2.3  &  88   &  4.4    &   &  5.0   &  2.3  &  91   &  9.1  \\ 
G09.DR1.47   &  4.8   &  2.9   &  113  &  7.1   &   &  5.4  &  2.7  &  187  &  47.0   &   &  5.7   &  2.7  &  379  &  189.9  \\ 
G09.DR1.49   &  2.5   &  2.4   &  220  &  0.1   &   &  4.4  &  2.8  &  160  &  4.0    &   &  4.8   &  2.8  &  113  &  7.1  \\ 
G09.DR1.53   &  4.4   &  2.8   &  160  &  4.0   &   &  5.0  &  2.5  &  110  &  11.0   &   &  5.3   &  2.6  &  187  &  37.3  \\ 
G09.DR1.56   &  3.5   &  2.5   &  221  &  0.7   &   &  4.5  &  2.6  &  103  &  3.3    &   &  4.8   &  2.6  &  92   &  5.8  \\ 
G09.DR1.60   &  4.4   &  2.6   &  132  &  3.3   &   &  4.9  &  2.2  &  73   &  5.8    &   &  5.2   &  2.2  &  78   &  12.4  \\ 
G09.DR1.61   &  3.6   &  2.7   &  221  &  0.9   &   &  4.7  &  2.8  &  124  &  6.2    &   &  5.0   &  2.8  &  140  &  14.0  \\ 
G09.DR1.62   &  4.4   &  3.0   &  197  &  4.9   &   &  5.2  &  2.9  &  140  &  22.2   &   &  5.5   &  2.9  &  568  &  179.6  \\ 
G09.DR1.72   &  4.1   &  2.6   &  172  &  2.2   &   &  4.8  &  2.4  &  80   &  5.0    &   &  5.1   &  2.4  &  91   &  11.4  \\ 
G09.DR1.80   &  2.0   &  2.0   &  263  &  0.0   &   &  3.9  &  3.5  &  368  &  2.9    &   &  4.5   &  3.7  &  270  &  8.5  \\ 
G09.DR1.85   &  2.8   &  2.3   &  204  &  0.1   &   &  4.3  &  2.6  &  132  &  2.6    &   &  4.7   &  2.7  &  103  &  5.2  \\ 
G09.DR1.87   &  5.1   &  2.9   &  140  &  17.6  &   &  5.7  &  2.7  &  379  &  189.9  &   &  5.8   &  2.7  &  720  &  454.3  \\ 
G09.DR1.99   &  4.1   &  3.0   &  249  &  3.1   &   &  5.1  &  2.9  &  140  &  17.6   &   &  5.5   &  3.0  &  799  &  252.7  \\ 
G09.DR1.113  &  4.7   &  3.0   &  157  &  7.9   &   &  5.4  &  2.8  &  255  &  64.1   &   &  5.6   &  2.8  &  568  &  226.1  \\ 
G09.DR1.125  &  3.3   &  2.4   &  191  &  0.4   &   &  4.4  &  2.7  &  132  &  3.3    &   &  4.8   &  2.7  &  92   &  5.8  \\ 
G09.DR1.127  &  5.1   &  3.1   &  186  &  23.4  &   &  5.7  &  2.9  &  568  &  284.7  &   &  5.9   &  2.9  &  1190 &  945.3  \\ 
G09.DR1.159  &  2.0   &  2.0   &  263  &  0.0   &   &  4.0  &  2.8  &  209  &  2.1    &   &  4.5   &  2.9  &  124  &  3.9  \\ 
G09.DR1.179  &  4.9   &  3.0   &  147  &  11.7  &   &  5.5  &  2.8  &  568  &  179.6  &   &  5.7   &  2.8  &  568  &  284.7 \\ 
G09.DR1.185  &  3.8   &  2.3   &  164  &  1.0   &   &  4.5  &  2.0  &  79   &  2.5    &   &  4.8   &  2.1  &  73   &  4.6  \\ 
G09.DR1.276  &  4.6   &  3.1   &  157  &  6.3   &   &  5.4  &  2.9  &  255  &  64.1   &   &  5.7   &  2.9  &  568  &  284.7  \\ 
G09.DR1.294  &  4.5   &  2.8   &  124  &  3.9   &   &  5.2  &  2.6  &  110  &  17.4   &   &  5.5   &  2.6  &  379  &  119.9  \\ 
G09.DR1.328  &  --    &  --    &  --   &  --    &   &  --   &  --   &  --   &  --     &   &  --    &  --   &  --   &  --  \\ 
\hline
\end{tabular}
\end{center}
\end{table*}

\subsection{Adjustments to observed quantities}\label{sec:adjustments}

In order to draw a direct comparison between our observations and the PDR model of \citet{kaufman1999,kaufman2006}, we must first make several adjustments to the observed \cii \ and CO line emission. 

Firstly, the \cii \ emission originates from both ionised and neutral gas, owing to carbon's lower ionization potential (11.26 eV) with respect to hydrogen (13.6 eV). Because the PDR models consider that the \cii \ emission originates purely from the neutral gas, we must therefore take into account the fraction of \cii \ emission arising from the ionised gas. A direct method to determine this fraction is by comparing ratios of the \cii ~158~$\mu$m and the \nii \ 122, 205 $\mu$m fine-structure emission lines, particularly the \cii /\nii 205 and \nii 122/\nii 205 ratios. The \nii 122/\nii 205 ratio is a sensitive probe of the ionised gas density in \hii \ regions, since the nitrogen ionization potential (14.5 eV) is greater than that of neutral hydrogen (13.6 eV). Because the \cii \ and \nii \ 205 $\mu$m lines have very similar critical densities for collisional excitation by electrons (46 and 44 cm$^{-3}$ at $T_{e}$ = 8000 K, respectively), the \cii /\nii 205 line ratios are mainly dependent on the relative abundances of C and N in the \hii regions. The ionised gas density can thus be inferred from the theoretical \nii 122/\nii 205 ratio, and used to predict the theoretical \cii /\nii 122 ratio arising from the ionised gas and subsequently estimate the neutral gas contribution to the \cii \ emission (for details see \citealp{oberst2006,oberst2011}). However, the lack of observations targeting these \nii \ transitions for our sample means we cannot exploit this method in this work. Instead, we must adopt the correction factors obtained from similar previous studies. 

Using the direct method, \citet{oberst2006} found that $\sim$73\% of the observed \cii \ line emission of the star-forming Carina nebula in the Galaxy arises from neutral gas in PDRs. Modelling of \hii and PDR regions of starburst galaxies NGC 253 and M82 (\citealp{carral1994}; \citealp{lord1996}; \citealp{colbert1999}) has also shown that PDRs account for $\sim$70\% of the \cii \ emission, with similar results (70 -- 85\%) reported by \citet{kramer2005} in their study of M51 and M83. Based on this evidence, numerous studies of higher redshift ($z\sim$ 1--6) IR-bright galaxies thus adopt a 0.7 correction factor to account for the ionised gas contribution to the \cii \  emission (see e.g. \citealp{stacey2010a}, \citealp{hailey2010}, \citealp{gullberg2015}). Since our sample comprises luminous (\lfir $\sim 10^{10}$--$10^{11}$~$L_{\odot}$), actively star-forming (SFR $\approx 40~\mathrm{M}_{\odot}\,\mathrm{yr}^{-1}$) main-sequence galaxies (see \citealp{villanuevaprep}), we also primarily assume $\sim$70\% of the observed \cii \ line emission (hereafter \ciipdrhi ) arises from PDRs. We note, however, the broader range of values found in the literature: the survey of \citet{malhotra2001}, for example, found that about 50\% of the observed \cii \ emission in their sample of galaxies originated in PDRs when using the \cii /\nii 205 emission combined with the \nii 122/\nii 205 ratio of our Galaxy. Spatially-resolved studies with the \textit{Herschel} Very Nearby Galaxies Survey have also demonstrated the varying distributions of neutral and ionised gas within local galaxies using the method of \citet{oberst2006,oberst2011} with both direct (e.g. \citealp{parkin2013,parkin2014}; \citealp{hughes2015}) and inferred (\citealp{hughes2015}) measurements of the \nii \ fine-structure lines. Within M51, the average ionised gas contribution to the \cii \ emission is $\sim$50\% in the spiral arm and interarm regions and reaches upto 80\% in the nucleus (\citealp{parkin2013}). In NGC 891, although the diffuse neutral component dominates the \cii \ emission in extraplanar regions, regions within the disk have up to 50\% of the \cii \ emission originating from ionised gas (\citealp{hughes2015}). Therefore, whilst at present it is still unclear what fraction of the \cii \ line emission arises from the ionized medium in galaxies, and a constant global correction will clearly not be appropriate in all sources, we choose to also examine the results of the PDR modelling when we assume that $\sim$50\% of the observed \cii \ line emission (hereafter \ciipdrlo ) comes from PDRs for all galaxies.

Secondly, we apply a correction to the \co \ emission to account for the likely case that this line becomes optically thick in dense star-forming regions much faster than the \cii \ line or the total infrared flux (see e.g. \citealp{stacey1983}; \citealp*{tielens1985}). The PDR infinite plane slab experiences an incident radiation field from one side, whereas the ensemble of clouds in extragalactic sources falling within the ALMA beam will not be orientated with their irradiated side facing towards our line of sight. We may therefore observe all the optically thin \cii \ line and FIR emission but overlook some of the optically thick \co \ line emission that escapes from those clouds with their irradiated sides orientated in the opposite direction from us. Previous studies make the assumption that we only observe about half of the total \co \ emission from all PDRs and thus multiply their observed \co \ emission by a factor of two. In the similar case of comparing the \cii \ 158 and \oi ~63~$\mu$m, \citet{stacey2010b} reason that such geometric issues should be accounted for by as much as a factor of four for high optical depth in a spherical cloud geometry. Here, we increase our observed \co \ emission by a factor of two for all galaxies, but keep in mind this may be a conservative correction. We refer to the combination of \ciipdrhi \ and 2$\times$CO as `standard adjustments' throughout the following text.

Finally, we note that some previous studies also include a correction to the total infrared flux from extragalactic sources, arguing that because the PDR model assumes the \lfir \ emission originates purely from the front side of the cloud, the observations must be reduced by a factor of two in order to account for the optically thin infrared continuum flux emitting not just towards the observer but from both sides of the PDR slab (see e.g. \citealp{parkin2013}; \citealp{hughes2015}), and hence make the \lfir \ emission directly equivalent to the bolometric far-infrared flux of the PDR model (see \citealp{kaufman1999}). We do not apply this correction here, choosing instead to follow the methodology of \citet{stacey2010a} and \citet{gullberg2015} in order to facilitate a comparison with these high-$z$ studies. As such, the possible contamination from an unknown fraction of \lfir \ that arises from ionised gas remains the main uncertainty in the FIR emission.

\subsection{Insights from the \lcii -- \lco \ diagnostic diagram}\label{sec:insights}

Following these adjustments, we now return to \hyperref[fig:pdrdiag]{Fig.~\ref*{fig:pdrdiag}} to examine their effects on the observations in the \lcii \ versus \lco \ parameter space. Focussing first on our \ciipdrhi -based correction (\hyperref[fig:pdrdiag]{Fig.~\ref*{fig:pdrdiag}}, semi-filled red circles),  we see that the comparison of these ratios to the contours on the diagnostic diagram indicates the majority of the galaxies have hydrogen nuclei densities in the range of $4 < \log n/\mathrm{cm}^{3} < 5$, and typically experience incident FUV radiation fields with strengths varying between $\log G_{0} \approx 2.0$ up to 3.0, though this increases towards $\log G_{0} = 3.5$ for some objects. Those galaxies previously falling outside of the \lcii/\lfir \ versus \lco/\lfir \ parameter space defined by the PDR model now lie within these same broad parameter ranges, although they appear to contain the least dense gas ($\log n/\mathrm{cm}^{3} \sim 3.5$) in comparison to the rest of the sample. When we consider the effect of the \ciipdrlo -based correction (\hyperref[fig:pdrdiag]{Fig.~\ref*{fig:pdrdiag}}, open red circles), we find that the expected overall shift of the data points to lower values of \lcii/\lfir \ translates into a slight increase in hydrogen nuclei densities, with an approximate range of $4.25 < \log n/\mathrm{cm}^{3} < 5.5$. In this region of the model parameter space, the FUV radiation field strengths depend primarily on the \lco/\lfir \ ratio and so there is no significant difference in our $G_{0}$ values when adopting either \ciipdrhi \ or \ciipdrlo \ as the diagnostic quantity. 

Despite the high quality of our ALMA \co \ and \textit{Herschel}-PACS \cii \ line observations, it is still difficult to accurately constrain the PDR model parameters using just two lines -- one ratio -- as diagnostics of the gas conditions due to degeneracies between the line ratio and the model parameter space. With access to more observations of far-infrared fine-structure lines (e.g. \oi ~63 or 145~$\mu$m) or lines arising from higher-J transitions of the CO molecule, it would be possible to more accurately determine the corresponding best-fit $n$ and $G_{0}$ values from the model by fitting the line ratios (using e.g. the PDRT online). Lacking this option, we nevertheless attempt to quantify the mean physical conditions in each galaxy by comparing the observations to the model parameter space. We use a KDTree search coupled with a $\chi^{2}$ minimisation technique to find the closest contour of the models to the \cii /\co \ observations and assign the corresponding $n$ and $G_{0}$ as the `best-fitting' values. We estimate the errors on the best-fitting model parameters via a bootstrap technique. After the best-fit $n$ and $G_{0}$ values are determined, 1000 new sets of data points are created by generating random flux values from within the errorbars of the observed or adjusted \cii , \co \ and FIR flux measurements. Each new dataset is then refit to find the alternative best-fitting parameter values. We then calculate the 68\% interval in both the upper and lower parameter distributions and set these intervals as the new upper and lower limits. The differences in the parameter values of the original best-fit solution and the extreme values generated from the bootstrap technique are taken as the uncertainties in the best-fit PDR model parameters. We stress that although this technique is satisfactory for assigning approximate PDR parameter values and their uncertainties based on the diagnostic diagram in \hyperref[fig:pdrdiag]{Fig.~\ref*{fig:pdrdiag}}, yielding typical errors of $\sim$0.1--0.3 dex in $\log n$ and $\sim$0.1 dex in $\log G_{0}$, a robust quantitative analysis remains limited by uncertainties in the coefficients used to adjust the observations (that are not considered with the bootstrap technique), the degeneracies in the model parameter space, and the fact we are treating each complete galaxy as a single PDR component. We discuss each of these issues in detail later.  

The results of our model fitting considering both the original and adjusted observations are presented in \hyperref[tab:params]{Table~\ref*{tab:params}}. For each case, the parameter ranges of course remain as discussed above. The unadjusted observations yield a mean incident field strength of 2.63$\pm$0.32 and density of 4.00$\pm$0.95 in logarithmic space. Using these best-fit $G_{0}$ and $n$ values, we predict the temperatures of the hot surface layer of the PDR atomic gas that faces towards the FUV source (the \emph{surface} temperature; see Fig. 1 in \citealp{kaufman1999}), which range from approximately 100 to 260~K with a mean temperature of 181 K. The majority of the PDR gas will be much cooler than this surface temperature. Equating the gas pressure to the product of the density and temperature, i.e. $P = nT$ with units of K~cm$^{-3}$, as in \citet{malhotra2001}, yields a range of $2\times 10^{4} < P < 24 \times 10^{6}$~K~cm$^{-3}$ with a mean of 5$\times 10^{6}$~K~cm$^{-3}$. On the other hand, the physical conditions of the gas traced with \ciipdrhi \ and CO adjusted upwards by a factor of two are on average denser ($\log n =$4.52$\pm$0.49), warmer ($T$=207$\pm$68~K) and more pressurised (47$\times 10^{6}$~K~cm$^{-3}$) than those results from the unadjusted observations, despite experiencing radiation field strengths of the same order of magnitude ($\log G_{0} =$2.25$\pm$0.25). Similar results are found when using \ciipdrlo \ as the PDR diagnostic (c.f. \hyperref[tab:params]{Table~\ref*{tab:params}}).

\section{Discussion}\label{sec:discussion}

To summarise the results from a comparison of our adjusted observations to the \citet{kaufman1999,kaufman2006} PDR model predictions, we find that, when using standard adjustments to observable quantities, the majority of the galaxies in our sample have hydrogen nuclei with densities ranging from 4$< \log n/\mathrm{cm}^{3} < 5.5$ with a mean of $\log n/\mathrm{cm}^{3} \sim$ 4.9, and experience an incident FUV radiation field with a strength between $2 < \log G_0 < 3$ normalised to the \citet{habing1968} Field (see \hyperref[tab:params]{Table~\ref*{tab:params}}). These correspond to mean gas temperatures of 207 K with pressures of 47$\times 10^{6}$~K~cm$^{-3}$. However, these results rely heavily on certain assumptions made in our analysis. Before comparing our results to those of previous studies in the literature, we first discuss the effects of these assumptions on our results.

\subsection{Assumptions and uncertainties}\label{sec:unc}

Firstly, there are uncertainties in the PDR parameters associated with our choice of PDR model. \citet{rollig2007} performed a detailed comparison of PDR models to examine the effects on the physical properties and chemical structures of the model clouds when using different PDR model codes in the literature. An important feature of a PDR model is the adopted geometry of the PDR region; the plane-parallel geometry of the \citet{kaufman1999,kaufman2006} model is a first order approximation and it is likely that a spherical model could be a more realistic approximation of the complex geometry of the PDR regions. Whilst the benchmarking exercise demonstrated that resulting trends in physical parameters are consistent between the participating codes, they warn that discrepancies remain between the observables computed with different codes (e.g. the atomic fine-structure line intensities) and that these uncertainties should be kept in mind when comparing PDR model results to observations in order to constrain physical parameters, such as density, temperature and radiation field strength. 

In applying the PDR model of \citet{kaufman1999,kaufman2006}, we adjusted the observations by (i) correcting the \cii ~158~$\mu$m line emission to remove the contribution to the emission arising from diffuse ionised gas, and (ii) increasing the \co \ emission by a factor of two to account for the likely case that this line becomes optically thick in dense star-forming regions. We assume these adjustments are correct to the first order for facilitating a proper comparison of our observations and the model. As we previously mention (\hyperref[sec:adjustments]{Sec.~\ref*{sec:adjustments}}), without an observational constraint of the fraction of the \cii \ line emission arising from the ionized medium in each of our galaxies, we must use evidence in the literature for global (e.g. \citealp{carral1994}; \citealp{lord1996}; \citealp{colbert1999}; \citealp{malhotra2001}) and resolved observations of galaxies (e.g. \citealp{parkin2013}; \citealp{parkin2014}; \citealp{hughes2015}) and assume that $\sim$50--70\% of the observed \cii \ line emission comes from PDRs for all galaxies. Although this is a standard assumption (see e.g. \citealp{stacey2010a}, \citealp{hailey2010}, \citealp{gullberg2015}), applying a constant global correction will not be appropriate for all galaxies and \lcii \ calculated from \ciipdrlo \ should be considered a lower limit (thus an upper limit in gas density). In the case of the correction to the \co \ emission, there remains a possibility that the line intensity should be further corrected by a factor of two (i.e. a factor of four in total) for high optical depth in a spherical cloud geometry, in contrast to the infinite slab geometry of the PDR model \citep{stacey2010a}. This correction would shift our galaxies to very low FUV radiation field strengths ($\log G_0 < 1.0$) not typically seen in nearby or higher redshift galaxies. A similar effect would be seen if optical depth effects become increasingly important in the handful of high-inclination systems in our sample (see e.g. NGC~891; \citealp{hughes2015}).

\begin{figure}
\begin{center}
\includegraphics[width=0.99\columnwidth]{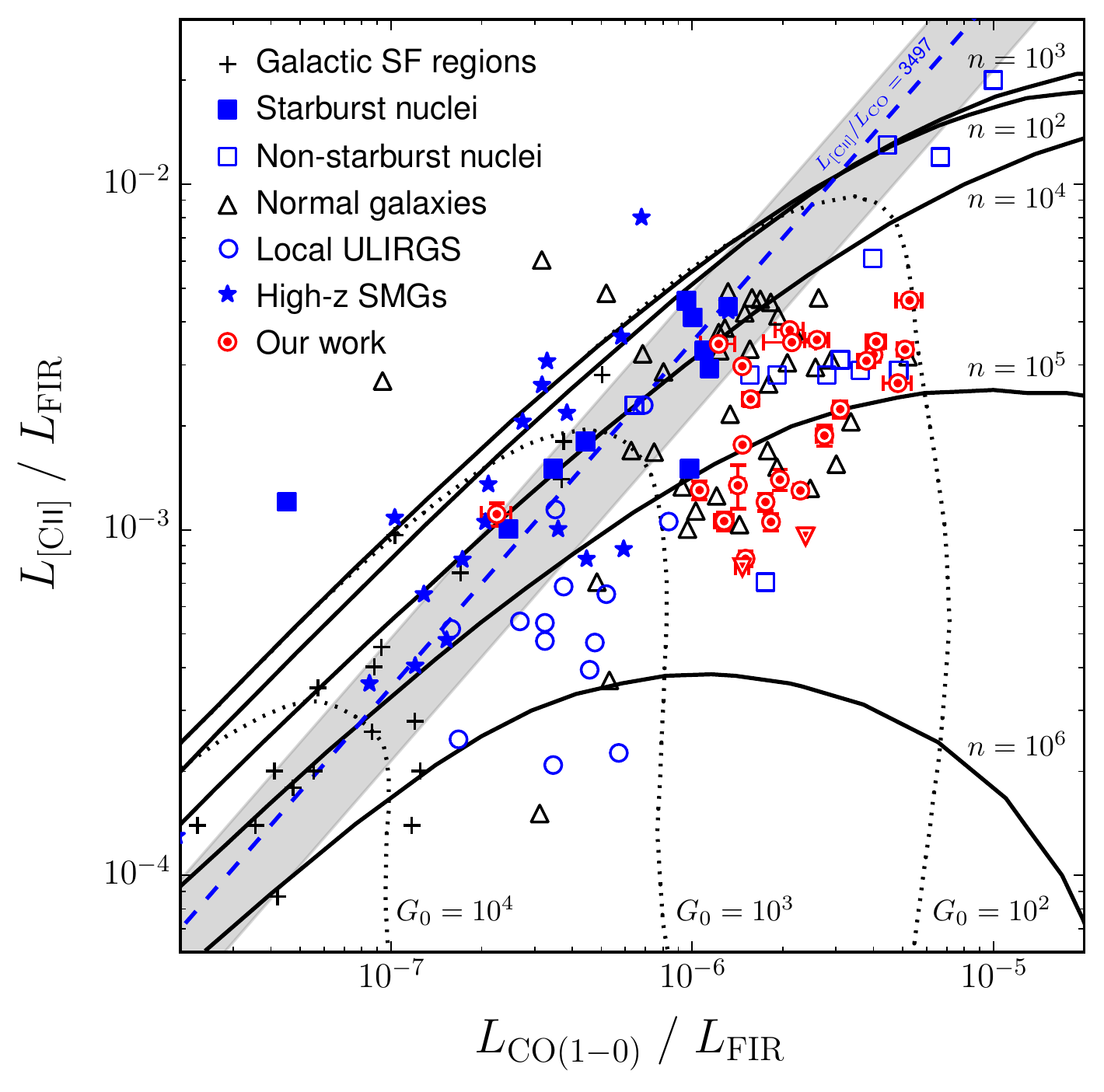}
\end{center}
\vspace{-0.5cm}
\caption{The diagnostic diagram of the \lcii/\lfir \ versus \lco/\lfir \ ratios (see also \hyperref[fig:pdrdiag]{Fig.~\ref*{fig:pdrdiag}} caption). We superimpose our adjusted observations, assuming 70\% of the total \cii \ emission arises from PDR regions (red semi-open circles), onto the grid of constant hydrogen nuclei density, log $n$ (black solid lines), and FUV radiation field strength, log $G_{0}$ (black dotted lines), determined from the PDR model of \citet{kaufman1999,kaufman2006}. Downward triangles represent upper limits. The comparative galaxy samples of \citet{hailey2010}, where this figure is adapted from, are also shown with the corresponding symbols in the legend. The high-$z$ submillimetre galaxies (blue stars) are from \citealp{gullberg2015}. The grey shaded region denotes the 1-$\sigma$ spread in the mean observed \lcii /\lco \ ratio (blue dashed line).}\label{fig:pdrdiaglit}
\end{figure}

Another possible correction would be to reduce the FIR emission by a factor of two, in order to account for the optically thin continuum flux emitting not just towards the observer but from both sides of the PDR slab (see \hyperref[sec:adjustments]{Sec.~\ref*{sec:adjustments}}), since the PDR model assumes the \lfir \ emission originates purely from the front side of the cloud. In addition, it may also be necessary to further reduce the TIR emission to account for continuum emission from other non-PDR sources, such as \hii regions and ionised gas (see e.g. \citealp{croxall2012}, who assume 70\% of the TIR emission originates from cool diffuse gas). Whilst we do not make such adjustments in this work, any reduction to \lfir \ would shift our \ciipdrhi - and \ciipdrlo -based adjusted observations upwards and to the right in the \lcii/\lfir \ versus \lco/\lfir \ parameter space in the \hyperref[fig:pdrdiag]{Fig.~\ref*{fig:pdrdiag}} diagnostic diagram (and also in \hyperref[fig:pdrdiaglit]{Fig.~\ref*{fig:pdrdiaglit}}), shifting $n$ and $G_0$ to lower densities ($\log n/\mathrm{cm}^{3} \approx 4.0$) and lower FUV field strengths ($\log G_0 \approx 1.5$). 

We further caution that additional shifts in the diagnostic diagrams may arise due to uncertainties with the ALMA and \textit{Herschel} flux calibration, and small, unknown offsets in the spatial coincidence of the \cii , \co \ and dust emission. At present, we expect such random shifts to be much smaller, secondary effects compared to the significant uncertainties in the observable quantities and model predictions discussed above. Although there are thus many uncertainties that apply to galaxies on an individual basis, we expect that the systematic adjustments we make are correct on average such that the dervied $G_0$ and $n$ values are representative. All of these limitations should be kept in mind throughout the following discussion.
   
\subsection{Comparison with previous studies}

As a first step in comparing the physical conditions derived for our sample to those in the literature, in \hyperref[fig:pdrdiaglit]{Fig.~\ref*{fig:pdrdiaglit}} we present the \lcii/\lfir \ versus \lco/\lfir \ diagnostic diagram with the contours of constant hydrogen nuclei density and FUV radiation field strength predicted by the \citet{kaufman1999,kaufman2006} PDR model (see \hyperref[fig:pdrdiag]{Fig.~\ref*{fig:pdrdiag}}), and superimpose our adjusted observations, assuming 70\% of the total \cii \ emission arises from PDR regions, together with the galaxy samples compared in Fig. 3 of \citet{hailey2010}, from where this figure is taken. To the first order, our galaxies fall in the region of the parameter space (i.e. $2 \leq \log G_0 \leq 3$ and $3 \leq \log n/\mathrm{cm}^{3} \leq 5$) predominantly populated by normal galaxies and non-starburst nuclei, in contrast to the Galactic star-forming regions and starburst nuclei that can exhibit higher radiation field strengths ($G_0>10^{3}$) and slightly higher densities ($n>10^{5}~\mathrm{cm}^{-3}$).

The gas properties we find in our sample are also consistent with other previous surveys of global, integrated observations not present in the figure of \citet{hailey2010}, although the hydrogen nuclei density tends to be higher on average than the literature values for nearby galaxies. For example, the \citet{malhotra2001} ISO survey uncovered densities of $2 \leq \log n/\mathrm{cm}^{3} \leq 4.5$, radiation field strengths of $2 \leq \log G_0 \leq 4.5$ and surface temperatures of $\sim$270--900~K, and similar results have since been uncovered in numerous other samples (c.f. Table 9 in \citealp{parkin2013}; see also e.g. \citealp{lebouteiller2012}). The detailed view afforded by spatially-resolved studies of nearby objects also indicates that our sample consists of normal star-forming galaxies. In particular, the Very Nearby Galaxy Survey (VNGS; P.~I.:~C.~D.~Wilson; see e.g. \citealp{parkin2013,parkin2014}; \citealp{schirm2014}; \citealp{hughes2014,hughes2016}), a \textit{Herschel} Guaranteed Time Key Project that aims to study the gas and dust in the ISM of a diverse sample of thirteen nearby ($D_\mathrm{L}<$90 Mpc) galaxies, has provided several spatially-resolved studies. In the case of M51 \citep{parkin2013}, the spiral arm and interarm regions exhibit hydrogen densities and FUV radiation field strengths of $2.75 \leq \log n \leq 3$ cm$^{-3}$ and $2.25 \leq \log G_0 \leq 2.5$, respectively, and typically reach higher ranges of $n$ and $G_{0}$ for both the central ($3 \leq \log n/\mathrm{cm}^{3} \leq 3.5$, $2.75 \leq \log G_0 \leq 3$) and nuclear ($3.75 \leq \log n/\mathrm{cm}^{3} \leq 4$, $3.25 \leq \log G_0 \leq 3.75$) regions. Comparable distributions in the physical conditions are also seen in the disks of Centaurus A \citep{parkin2014} and NGC 891 \citep{hughes2015}, though the high inclination of the latter system presents an additional challenge in their interpretation. The K\textsc{ingfish} team also report that the PDR gas in NGC 1097 and NGC 4559 has densities between 10$^{2.5}$ and 10$^{3}$ cm$^{-3}$ whereas $G_0$ ranges from 50 to 1000 when adopting conservative corrections for ionized gas (see Fig. 10 of \citealp{croxall2012}).

In the case of higher redshift studies, \citet{gullberg2015} present \cii \ observations of gravitationally lensed dusty star-forming galaxies in the redshift range $z\sim$2.1--5.7 discovered by the South Pole Telescope (SPT; \citealp{carlstrom2011}). A subsample also has \co \ observations (\citealp{aravena2016b}). Using the \lcii/\lfir \ versus \lco/\lfir \ diagnostic diagram (see their Fig.~12), the SPT sample observations suggest the radiation field strength and average gas density to be in the range $100<G_0<8000$ and $10^{2}<n<10^5$\,cm$^{-3}$ for unadjusted observations. Applying the adjustments to obtain 2$\times$CO and \ciipdrhi \ for these galaxies increases the gas density range by an order of magnitude. In addition to the SPT galaxies, \citet{gullberg2015} also assemble a high redshift ($2.33< z <6.42$) sample of galaxies in the literature with available measurements of the \cii ~158\,$\mu$m and \co \ line emission and FIR luminosities (see their Table B1 and accompanying references therein). On the \lcii/\lfir \ versus \lco/\lfir \ diagnostic diagram, these systems have observed ratios corresponding to field strengths of $2<\log G_0<4$ and densities of $2<\log n/\mathrm{cm}^{3}<6$, which translate to much denser PDR regions ($\sim$10$^{5}\mathrm{cm}^{-3}$) encountering similar field strengths if we apply the adjustments of 2$\times$CO and \ciipdrhi \ , as detailed above. In comparison, our sample galaxies therefore typically have lower density gas encountering weaker FUV fields than these higher redshift SPT systems.

\begin{figure*}
\begin{center}
\includegraphics[width=253.514052pt]{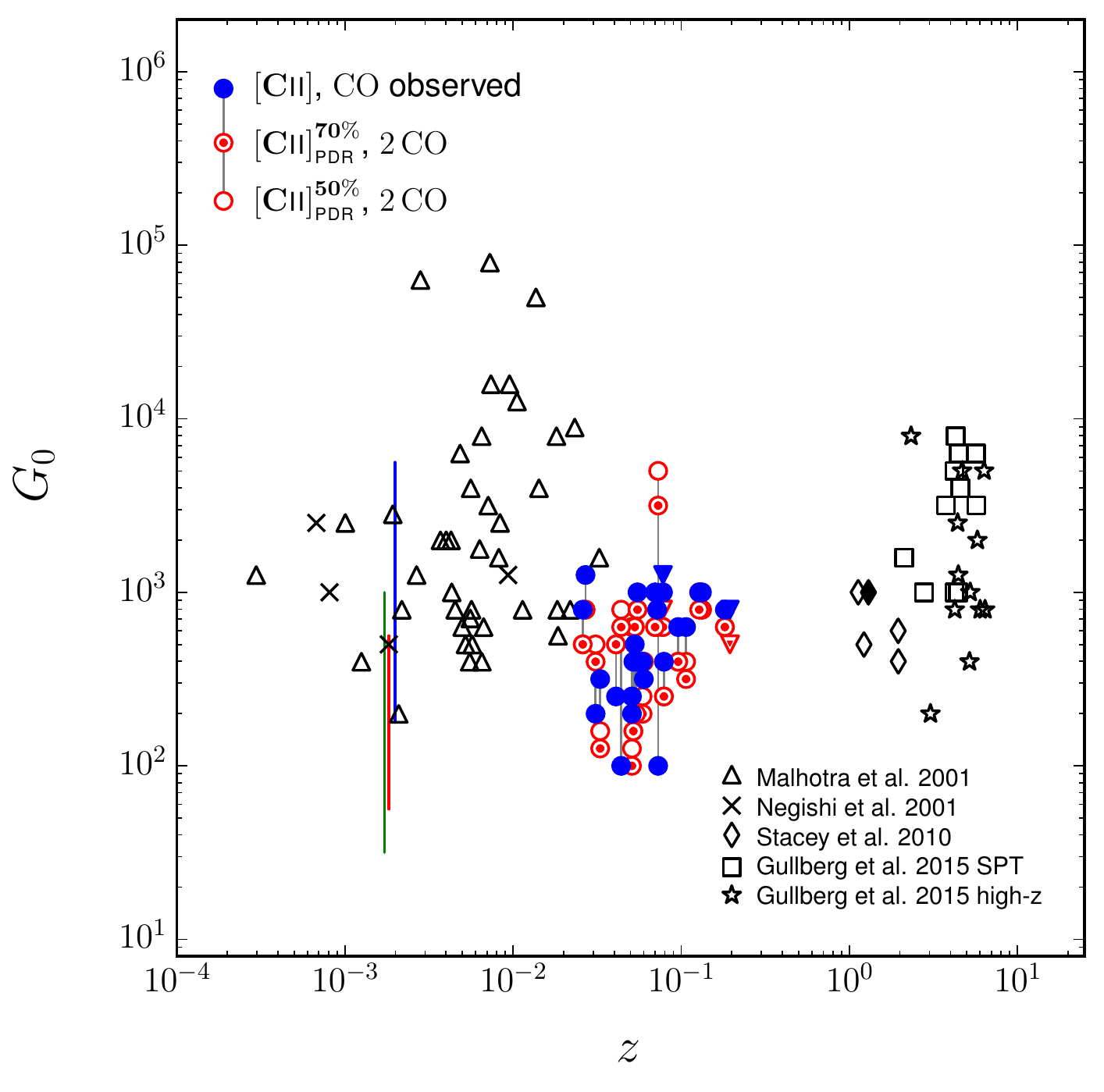}
\includegraphics[width=253.514052pt]{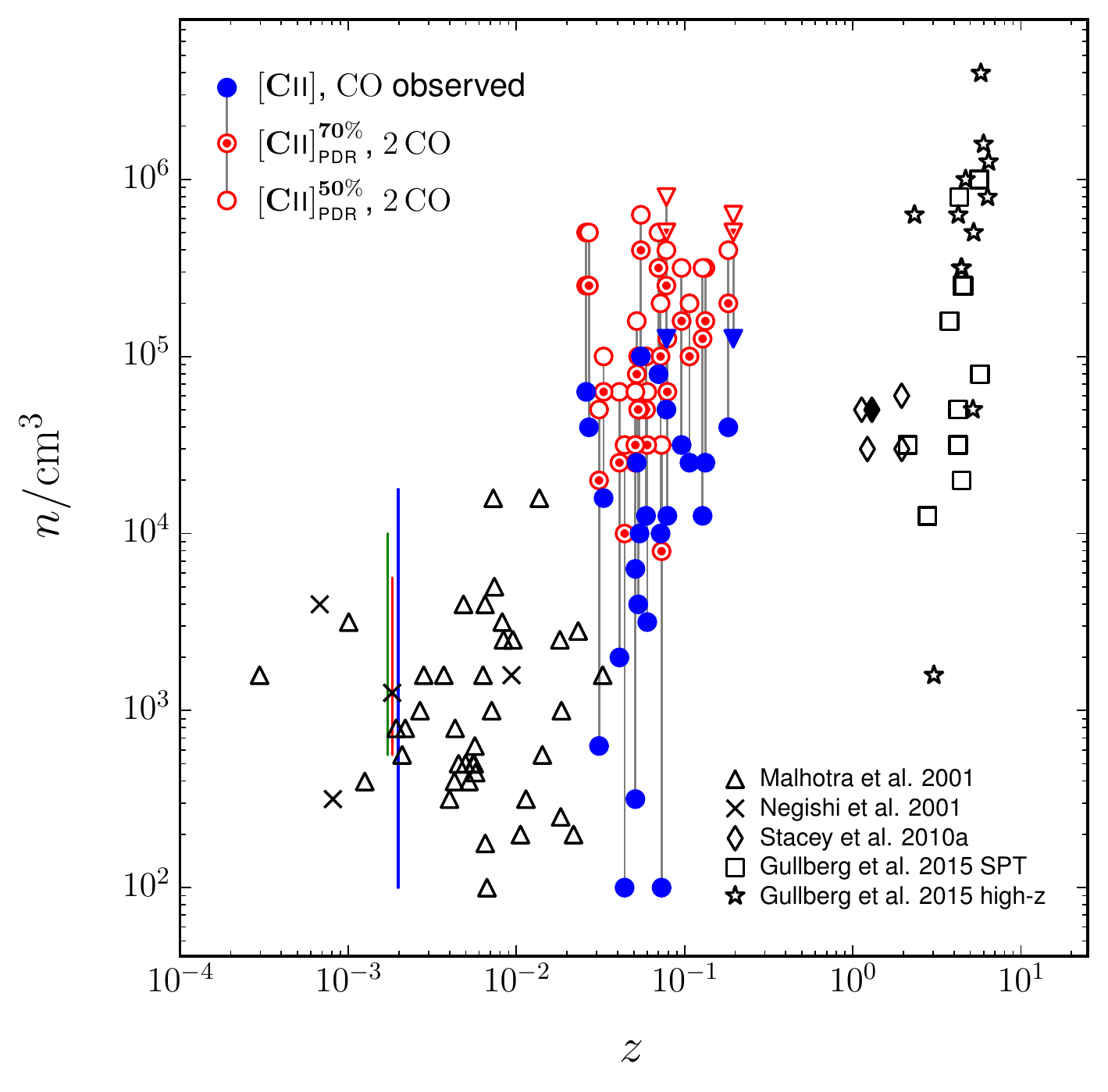}
\end{center}
\vspace{-0.5cm}
\caption{The average $G_{0}$ (left panel) and $n$ (right panel) plotted as a function of galaxy redshift. We compare the values derived from the unadjusted observations (blue solid circles) and the observations including the adjustments applied to the \cii \ and \co \ emission (see \hyperref[sec:adjustments]{Sec.~\ref*{sec:adjustments}}), assuming either 70\% (red semi-open circles) or 50\% (red open circles) of the total \cii \ emission arises from PDR regions. Downward triangles represent upper limits. Comparative galaxy samples of the literature are shown with the symbols specified in the legends, together with the ranges seen in spatially-resolved studies of M51 (blue line; \citealp{parkin2013}), Cen A (red line; \citealp{parkin2014}) and NGC 891 (green line; \citealp{hughes2015}). To facilitate a comparison, we apply the same standard adjustments (i.e. adopting \ciipdrhi \ and 2CO) to the high redshift ($z>1$) systems as those made for galaxies at lower redshift.}\label{fig:zparams}
\end{figure*}

\subsection{Is there a cosmic evolution of the physical conditions?}

As hinted in the previous section, there appear to be some offsets between our galaxies and those at low and high redshift, which may suggest an evolution in the average gas density of the PDR gas. An evolution of the mean gas density would be an interesting result, providing an explanation for the evolution in the star formation rate density of the Universe, $\rho_\mathrm{SFR}$ (see e.g. \citealp*{madau2014}, and references therein), whereby on average galaxies contain denser neutral gas at higher redshifts that can more readily be converted into stars than in their low-redshift counterparts. Given that the star formation activity in the Universe was significantly higher in the past and the FUV radiation field strength within galaxies will be determined by the star formation history and the initial mass function, with a small contribution from the cosmic FUV background, we expect to observe an evolution of $G_0$ with redshift. Additional constraints may also be gained from the evolution of the physical conditions; for example, following standard assumptions that the galaxy mass-to-light ratio ($M/L$) and the surface brightness ($L/R^{2}$) are constant, that $n$ is a constant fraction of the total galaxy mass $M$ such that $n \propto M/R^{3}$, and the FUV radiation field strength scales as $G_{0} \propto L$, we would expect that the quantity $G_{0} n^{2}$ is also constant as a function of redshift. A robust refute of this would therefore have important implications for e.g. the Tully-Fisher relation (\citealp{tullyfisher1977}), since recent studies show the relation exhibits weak/no evolution with redshift (see e.g. \citealp{miller2012}; \citealp{molinaprep}).

In \hyperref[fig:zparams]{Fig.~\ref*{fig:zparams}}, we present the $G_{0}$ and $n$ values for each galaxy plotted as a function of galaxy redshift. Whilst we primarily focus on our adjusted observations with adjustments based on the standard assumptions of \ciipdrhi \ optically-thick CO emission (i.e. increasing CO by a factor two), we also plot our original, unadjusted observations together with the adjusted observations via \ciipdrlo \ to demonstrate the possible ranges of our observations. We compare our data to as many literature values as possible under the condition that they are derived using the same PDR model, in order to avoid any complications arising from the aforementioned offsets between different PDR codes (e.g. \citealp{rollig2007}), and with similar adjustments to account for the fraction of \cii \ from ionised gas. Comparative galaxy samples of the literature are shown with varying symbols in the legends, together with the ranges seen in spatially-resolved studies of M51 (blue line; \citealp{parkin2013}), Cen A (red line; \citealp{parkin2014}) and NGC 891 (green line; \citealp{hughes2015}). 

Interestingly, the FUV radiation field strength does not appear to show any apparent overall evolution with redshift; the low redshift ($z<0.05$) sample of global integrated observations exhibit an average FUV radiation field strength of $\sim$10$^{3}$, with most galaxies lying within an order of magnitude above and below this value, and the high redshift ($z>1$) systems show a similar mean and distribution around $G_{0}\sim$10$^{3}$. Our sample typically lies below $\sim$10$^{3}$ regardless of whether or not we apply any adjustments to our observations. It is worth noting that to match the mean $G_{0}$ of our sample to the low and high redshift values would require a significant reduction ($\sim$50\%) in both the \cii \ and \co \ emission, whereas most of the uncertainties in our methodology discussed in \hyperref[sec:unc]{Sec.~\ref*{sec:unc}} would in fact lead to an increase of the emission. In contrast, the hydrogen nuclei density appears to show a significant redshift evolution. Although the mean density of both the integrated and spatially-resolved observations of galaxies at low redshift is $n\sim$10$^{3}$, our galaxies exhibit a higher mean density between $n\sim$10$^{4}$--10$^{5}$ depending on the adopted correction (see the right panel of \hyperref[fig:zparams]{Fig.~\ref*{fig:zparams}}), roughly equivalent to the density distribution of the high-$z$ systems, that suggests a strong evolution of the density up to $z=0.2$. We briefly note that $G_{0} n^{2}$ is not constant, contrary to our prediction, as we observe a similar variation with redshift that is driven by the variation in the gas density.

However, it is not possible at this stage to firmly conclude that this observed evolution is in fact physical and we must be cautious not to confuse clear observational biases with an evolution of the physical conditions. One bias may be introduced from differences in the samples at low and high redshift. The SPT sample comprises submillimetre-selected dusty star-forming galaxies, which are typically massive in both their stellar content (M$_{*}$ $\sim 10^{11} \mathrm{M}_{\odot}$) and gas content (M$_{\mathrm{gas}}$ $\sim 5\times10^{10}~\mathrm{M}_{\odot}$) and have star formation rates of the order of 1000$~\mathrm{M}_{\odot}\,\mathrm{yr}^{-1}$ (see \citealp{gullberg2015}, and references therein). In contrast, the corresponding key properties of our \textit{H}-ATLAS-based sample suggest our galaxies are normal star-forming systems (see \citealp{villanuevaprep}) and so may not be directly comparable to these high redshift systems that generally lie above the main sequence (see e.g. \citealp{elbaz2011}). The potential for misclassification of high-$z$ sources and contamination of the sample (e.g. from AGN hosts) only adds to these difficulties. Nevertheless, we observe an offset between the VALES sample and the low-redshift systems that do share similar properties and do appear comparable (e.g. \citealp{malhotra2001}; \citealp{negishi2001}; \citealp{parkin2013}), but this potential evidence for an evolution in the gas density is also undermined by a bias that may arise from using different suites of emission lines as PDR diagnostics. The results at low redshift are based on a whole suite of far-infrared fine structure lines (e.g. the \cii \ 158~$\mu$m, \oi \ 63 and 145~$\mu$m, \oiii \ 88~$\mu$m and \nii \ 122 and 205~$\mu$m,  lines) and derivative diagnostic ratios (e.g.  \ciioi \ or \ciioiltir ) that may be probing a different density regime of the PDR regions compared to the higher redshift observations that rely on only the \cii \ and \co \ line emission. Furthermore, we note that very nearby galaxy studies tend to be more hand-picked and less statistical in their selection, thus may not be appropriate for comparison to IR-selected high-z samples. The difference between the nearby galaxies and our sample is most likely owing to differences in methodology and/or sample selection and, for these reasons, we recommend our VALES sample as a better local reference for more distant galaxies when using the \cii \ and \co \ emission lines.

At present, we therefore cannot draw strong conclusions on whether the gas density evolves with redshift. Placing a more robust constraint on the evolution in the average $G_{0}$ and $n$ values would require additional gas diagnostics via observations of other far-infrared fine-structure lines in higher redshift systems. For example, the ratio of the two submillimetre C$^{0}$ transitions, \ci \ 370~$\mu$m/\ci \ 609~$\mu$m, is particularly useful for tracing the gas density for the following reasons (see e.g. \citealp{kaufman1999}): (1) a weak temperature dependence due to low upper-state energies for both transitions ($E_{609}/k \sim 24$ K, $E_{370}/k \sim 63$ K); (2) an insensitivity to the FUV radiation field strength in temperatures warm enough to excite both transitions; (3) similar critical densities ($n_{crit}($\ci ~$370~\mu \mathrm{m}) \sim 3\times 10^{2}~\mathrm{cm}^{-3}$ and $n_{crit}($\ci ~$609~\mu \mathrm{m}) \sim 2\times 10^{3}~\mathrm{cm}^{-3}  )$; and (4) an insensitivity to the C$^{0}$ abundance. For high values of the FUV radiation field strength per hydrogen nuclei density ($G_0/n$), the FUV heating of the gas occurs at $A_{V}<1$ and the \ci \ lines are generally weaker than the \cii \ and \oi \ lines, whereas C$^{0}$ lines tend to dominate gas cooling at small column densities in clouds with low values of $G_0/n$, i.e. $3\times 10^{-3}$. The \ci \ lines are therefore useful for determining the details of the chemistry and penetration of the incident FUV in photodissociation regions. Of course, the \nii ~122~and~205~$\mu$m fine-structure emission lines for constraining the contribution to the \cii \ emission from ionised gas would also be valuable.

\section{Conclusions}\label{sec:conclusions}

We have used new ALMA \co \ observations to study the physical conditions in the interstellar gas of a sample of 27 dusty main-sequence star-forming galaxies at $0.03<z<0.2$, bridging the gap between local and high-$z$ galaxy samples. Our sample was drawn from far-IR bright galaxies over $\sim$160 deg$^{2}$ in the \textit{Herschel} Astrophysical Terahertz Large Area Survey. We combined these observations with high-quality ancillary data, including \textit{Herschel}-PACS \cii \ 158~$\mu$m spectroscopy and far-infrared luminosities determined with photometry from \textit{Herschel} and other facilities. Our main results and conclusions are:
\\\\
\textbf{1.} The average line luminosity ratio of the \cii \ and \co \ detections is $3500\pm1200$ for our sample, in agreement with previous works. The observed line emission is also consistent with that of both local and high-$z$ galaxies. The observed \lcii / \lco ratios appear to arise from gas with $T_{\mathrm{ex,[C{\textsc{ii}}]}} > T_{\mathrm{ex,CO(1-0)}}$, where the \cii \ is optically thin and the CO is optically thick.
\\\\
\textbf{2.} Our sample covers the same distribution in the \lcii /\lfir \ and \lco /\lfir \ ratio as normal star-forming galaxies and non-starburst nuclei. 
\\\\
\textbf{3.} A comparison with the \lcii/\lfir \ versus \lco/\lfir \ diagnostic diagram in \hyperref[fig:pdrdiag]{Fig.~\ref*{fig:pdrdiag}} provides a first order estimate of the radiation field strength and average gas density found to be in the range of  $2 < \log G_0 < 3$ and $2 < \log (n/\mathrm{cm}^{3}) < 4.5$. Assuming standard adjustments increases the gas density by an order of magnitude. These values are consistent with previous surveys employing either integrated or spatially-resolved observations, although the hydrogen nuclei density tends to be higher on average than the literature values for observations of nearby galaxies.
\\\\
\textbf{4.} The average FUV radiation field strength appears constant up to redshift $z\sim6.4$, yet we find an increase of a factor of $\sim$100 in the neutral gas density as a function of redshift that persists regardless of various adjustments to our observable quantities. The apparent evolution most likely arises from a combination of observational biases rather than physical processes and we highlight the need to take a consistent methodology between different studies.  
\\\\
Of the many uncertainties in this work, the major limitation is the use of only two emission lines -- one ratio -- as diagnostics of the physical conditions of the gas. Future studies with ALMA could observe these galaxies using other fine-structure lines, particularly the two submillimeter C$^{0}$ transitions, \ci ~370~and~609~$\mu$m, the ratio of which is useful for tracing the gas density and determining the details of the chemistry and penetration of the incident FUV in photodissociation regions. The capability of ALMA to spatially resolve the \cii \ emission and other fine structure lines in galaxies at higher redshifts will in future provide a more detailed picture of whether or not the gas density evolves with cosmic time.

\section*{Acknowledgements}
We thank the anonymous referee for their clear and constructive report. TMH and EI acknowledge CONICYT/ALMA funding Program in Astronomy/PCI Project N$^\circ$:31140020. This paper makes use of the following ALMA data: ADS/JAO.ALMA\#2013.1.00530.S. MA acknowledges partial support from FONDECYT through grant 1140099. ALMA is a partnership of ESO (representing its member states), NSF (USA) and NINS (Japan), together with NRC (Canada), NSC and ASIAA (Taiwan), and KASI (Republic of Korea), in cooperation with the Republic of Chile. The Joint ALMA Observatory is operated by ESO, AUI/NRAO and NAOJ. The \textit{Herschel}-ATLAS is a project with \textit{Herschel}, which is an ESA space observatory with science instruments provided by European-led Principal Investigator consortia and with important  participation from NASA. The H-ATLAS website is \url{http://www.h-atlas.org/}. PACS has been developed by a consortium of institutes led by MPE (Germany) and including UVIE (Austria); KU Leuven, CSL, IMEC (Belgium); CEA, LAM (France); MPIA (Germany); INAF-IFSI/OAA/OAP/OAT, LENS, SISSA (Italy); IAC (Spain). This development has been supported by the funding agencies BMVIT (Austria), ESA-PRODEX (Belgium), CEA/CNES (France), DLR (Germany), ASI/INAF (Italy), and CICYT/MCYT (Spain). SPIRE has been developed by a consortium of institutes led by Cardiff University (UK) and including Univ. Lethbridge (Canada); NAOC (China); CEA, LAM (France); IFSI, Univ. Padua (Italy); IAC (Spain); Stockholm Observatory (Sweden); Imperial College London, RAL, UCL-MSSL, UKATC, Univ. Sussex (UK); and Caltech, JPL, NHSC, Univ. Colorado (USA). This development has been supported by national funding agencies: CSA (Canada); NAOC (China); CEA, CNES, CNRS (France); ASI (Italy); MCINN (Spain); SNSB (Sweden); STFC, UKSA (UK); and NASA (USA). This research has made use of the NASA/IPAC Extragalactic Database (NED) which is operated by the Jet Propulsion Laboratory, California Institute of Technology, under contract with the NASA (USA).

\bibliography{aa29588-16}

\end{document}